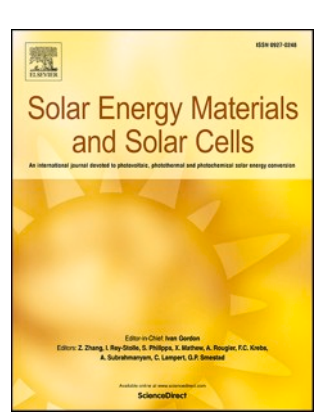

# Unraveling the defect landscape of wide-bandgap perovskites from electrical and photoelectrical characterization of thin films and solar cells

L. Kopprio [a,b,c,*], J. Caram [d,e,f], S. Le Gall [a,b,c,**], F. Ventosinos [d,e,g,***], L. Gil-Escrig [g], H.J. Bolink [g], J. Alvarez [a,b], C. Longeaud [a,b], J.-P. Kleider [a,b,c,****], J. Schmidt [d,e]

[a] Université Paris-Saclay, CentraleSupélec, CNRS, Laboratoire de Génie Electrique et Electronique de Paris, Gif-sur-Yvette, 91192, France
[b] Sorbonne Université, CNRS, Laboratoire de Génie Electrique et Electronique de Paris, Paris, 75252, France
[c] Institut Photovoltaïque d'Ile-de-France (IPVF), 18 Boulevard Thomas Gobert, Palaiseau, 91120, France
[d] Instituto de Física del Litoral (IFIS-Litoral), CONICET-UNL, Güemes 3450, Santa Fe, S3000GLN, Argentina
[e] Facultad de Ingeniería Química, UNL, Santiago del Estero 2829, Santa Fe, S3000AOM, Argentina
[f] Centro de Investigación de Métodos Computaciones (CIMEC), CONICET-UNL, Santa Fe, S3000, Argentina
[g] Instituto de Ciencia Molecular, Universidad de Valencia, C/Catedrático J. Beltrán 2, Paterna, 46980, Spain

ABSTRACT

Understanding and controlling defect states in halide perovskites is critical to advancing their performance in solar cells, yet their complex defect landscape remains elusive. Charged defects in perovskites can migrate under an applied electric field, complicating their characterization by conventional approaches. Here, we integrate current-voltage (IV) and thermal admittance spectroscopy (TAS) with lateral photocurrent methods, including thermal steady-state photocurrent (SSPC) and steady-state photocarrier grating (SSPG), to probe the kinetic and electrical properties of defects in thin films of vacuum-deposited $FA_{0.7}Cs_{0.3}Pb(I_{0.9}Br_{0.1})_3$ perovskite. The experimental results are interpreted with advanced numerical simulations to account not only for the energy positions of defects in the bandgap but also for their mobilities. The low activation energies observed in the capacitance steps rule out free-carrier trapping and emission as their origin, pointing instead to charged-defect (or ionic) migration. We estimate the free-carrier mobilities and the defect distribution inside the bandgap, along with their capture coefficients. Our results reveal exponential bandtail states arising from dynamic lattice disorder and identify a Gaussian-like defect distribution 0.21 eV from the band edge, which dominates recombination. Donors and acceptors are present at nearly equal concentrations ($\sim 2 \times 10^{18}$ $cm^{-3}$). The mobile species responsible for the capacitance steps is one of the dopants, exhibiting an average mobility of $10^{-8}$ $cm^2 V^{-1} s^{-1}$ at 300 K with a thermal activation energy of around 0.34 eV.

## 1. Introduction

Lead halide perovskites (LHPs) with the general formula A-Pb-$X_3$, where A represents monovalent cations [methylammonium (MA), formamidinium (FA), cesium (Cs)] and X denotes halide anions (Cl, I, Br), have emerged as leading materials for next-generation solar cells, light-emitting diodes, and photodetectors, due to their exceptional optoelectronic properties, solution processability, and tunable band gaps. Unlike classical semiconductors such as silicon and gallium arsenide, where point defects create deep traps that significantly reduce carrier lifetimes, LHPs predominantly host shallow-level defects that have less impact on recombination. This defect resilience arises from the antibonding coupling between lead 6s and halide 5p orbitals, and the high ionicity of the crystal lattice, which usually localizes defect states near the band edges. Despite this inherent tolerance, surface and grain boundary defects remain critical performance limiters. In thin-film devices, these defects act as centers for non-radiative recombination, reducing photoluminescence quantum yields, open-circuit voltage, and device stability [1–4].

Most LHP research focuses on solution-processed thin films; however, vacuum deposition offers precise control over film thickness and composition, facilitates scale-up for large-area devices, and avoids solvent-related issues. Vacuum deposition also produces pinhole-free, uniform films with smooth morphology, advantageous for fabricating complex multilayer structures such as tandem solar cells [3,5,6].

* Corresponding author. Université Paris-Saclay, CentraleSupélec, CNRS, Laboratoire de Génie Electrique et Electronique de Paris, Gif-sur-Yvette, 91192, France.
** Corresponding author. Université Paris-Saclay, CentraleSupélec, CNRS, Laboratoire de Génie Electrique et Electronique de Paris, Gif-sur-Yvette, 91192, France.
*** Corresponding author. Instituto de Física del Litoral (IFIS-Litoral), CONICET-UNL, Güemes 3450, S3000GLN Santa Fe, Argentina.
**** Corresponding author. Université Paris-Saclay, CentraleSupélec, CNRS, Laboratoire de Génie Electrique et Electronique de Paris, Gif-sur-Yvette, 91192, France.
E-mail addresses: leonardokopprio@gmail.com (L. Kopprio), sylvain.le-gall@universite-paris-saclay.fr (S. Le Gall), fedevento@gmail.com (F. Ventosinos), jean-paul.kleider@centralesupelec.fr (J.-P. Kleider).

Among LHPs, mixed-cation and mixed-halide perovskites such as $FA_{0.7}Cs_{0.3}Pb(I_{0.9}Br_{0.1})_3$ offer improved phase stability, enhanced crystallinity, and superior device performance compared to their single-cation or single-halide counterparts [7–9]. The vacuum deposition of this perovskite could reach stabilized power outputs of 18% in single-junction devices [10]. The $FA_{1-x}Cs_xPb(I_{1-y}Br_y)_3$ perovskite group, if co-sublimed in a four-source reactor, can yield optical bandgaps within the 1.7-1.8 eV range without halide segregation, ideal for tandem top-cell absorbers [11,12].

Despite the remarkable success in increasing the efficiency and stability of Perovskite Solar Cells (PSCs), the underlying defect landscape of perovskites A-Pb-$X_3$ remains unclear. According to theoretical calculations [13], charged vacancies [such as $V_X^+$, $V_{Pb}^{2-}$, $V_A^-$, and $(V_{Pb}V_X)^-$] and interstitials (such as $X_i^-$ and $A_i^+$) are the most likely native point defects in lead halide perovskite. In thin film materials, these defects and others (such as uncoordinated $Pb^{2+}$) are concentrated in grain boundaries and interfaces [1], and could migrate in response to an applied electric field, which complicates their electrical characterization [14].

Another source of defects in thin film materials is the lattice disorder, which usually causes exponential tail states near the band edges. In thin films of halide perovskites, the dynamic contribution is typically much higher than the static one, even at low temperatures, and cannot be neglected, as it is often done in other thin film semiconductors, such as a-Si:H, ZnO and $CuIn_3Te_8$ [15–18]. The exponential energy dependence of these defect distributions and their temperature dependence are well established from the measurements of the sub-bandgap absorption coefficient, using Fourier Transform Photocurrent Spectroscopy (FTPS), Photothermal Deflection Spectroscopy (PDS), Photoluminescence (PL), and ultrasensitive External Quantum Efficiency (EQE) [19–23].

A straightforward method for characterizing the bulk properties of a thin film semiconductor is to perform experiments in the lateral direction. If the material is not strongly anisotropic, there is a clear correlation between the transport perpendicular to the surface (such as in a solar cell) and the lateral direction. Such is the usual case in thin films of halide perovskites [24,25]. At high temperatures ($T \geq 100$ K) and low generation rates ($G_r < 10^{22}$ cm$^{-3}$s$^{-1}$), the SRH mechanism dominates the recombination, allowing us to neglect the bimolecular and Auger mechanisms [26]. Consequently, lateral photoconductivity-based techniques in this regime are ideal for extracting the electrical properties of defects [3].

In this work, we combine standard Current-Voltage (IV) and Thermal Admittance Spectroscopy (TAS) in solar cells with lateral current-based techniques, such as the thermal Steady-State PhotoCurrent (SSPC) and the Steady-State Photocarrier Grating (SSPG), to estimate the kinetic and electrical properties of the defect states in thin films of vacuum-deposited $FA_{0.7}Cs_{0.3}Pb(I_{0.9}Br_{0.1})_3$ perovskite. Note that some studies [14,27] only focus on extracting the kinetic properties of defect states in halide perovskites, and others [4,28,29] only focus on extracting their electrical properties. The complete defect landscape of a perovskite material must include both types of properties.

Initially, we revisit TAS measurements in short-circuited PSCs. According to the literature [30–33], two mechanisms could produce the low-frequency capacitance steps observed experimentally: free-carrier trapping and emission by defects in the absorber and migration of charged defects in the perovskite layer. To identify the mechanism behind the capacitance signal, Refs. [31,33,34] suggest combining steady-state with transient measurements. Using numerical simulations, we show that it is not necessary to perform transient measurements when the thermal activation energy of the capacitance step is sufficiently small. In this case, the only way to reproduce the large steps of the experimental capacitance is through mobile charged defects (or ionic species).

Furthermore, we identify inconsistencies between standard drift-diffusion modeling and the previous analytical formulas [14,30] that relate the frequency position of the capacitance step to the kinetic properties of charged defects, such as the mobility. Using drift-diffusion modeling, we derive two new equations: one for symmetrical devices and the other for asymmetrical ones.

Then, the defect distribution in the bandgap is estimated from the current and photocurrent measurements performed in lateral devices with coplanar contacts. We follow a similar approach to Refs. [3,28], but we use a more realistic model that includes the bandtail states and uses Gaussian distributions for the defects. Finally, we estimate the perovskite electronic bandgap, permittivity, and their ionic mobility and concentration by fitting the solar cell IVs and TAS measurements.

This manuscript is organized as follows. Section 2 develops the theory of each applied technique. Section 3 presents the experimental methods, and Section 4 presents the experimental results and discussions. In Section 5, we extract a realistic set of material parameters that completely reproduce the experimental results with numerical simulations. The conclusion is presented in Section 6.

## 2. Theoretical developments

### 2.1. Thermal admittance spectroscopy (TAS) in halide perovskites

A general introduction to Thermal Admittance Spectroscopy (TAS) can be found in Section 1.A of the Supplementary Information (SI).

In high-quality short-circuited ($V_{DC} = 0$ V) PSCs [30,35], we observe a high-frequency plateau in the capacitance vs. frequency spectra, corresponding to the geometrical capacitance of the device, and an increase in capacitance as the signal frequency decreases. In principle, two mechanisms could increase the capacitance at low frequencies: free carrier trapping and emission by defects in the absorber, or migration (oscillation) of charged defects (ionic species) in the perovskite layer [30,31].

In Fig. 1, we present the simulated (Silvaco Atlas) dark capacitance of a typical (PIN structure) short-circuited PSC at 300 K. The basic parameters of the device were extracted from Table S1 of Ref. [32], which assumes undoped transport layers. In Fig. 1a, the low-frequency capacitance is produced by adding to the perovskite layer different concentrations of (immobile) defects at a fixed energy within the bandgap (0.9 eV from the valence band edge). In Fig. 1b, the low-frequency capacitance is produced by adding to the perovskite layer different concentrations of a completely ionized mobile dopant. In the first case, the low-frequency capacitance increases with the concentration of (immobile) defects (see Fig. 1a) and decreases as the defects energy approaches the band edges (see Fig. S1). In the second case (see Fig. 1b), the low-frequency capacitance increases with the concentration of mobile dopants, until it saturates to the value given by the series addition of the geometrical capacitances of the transport layers $\left(\frac{1}{C_{htl}} + \frac{1}{C_{etl}}\right)^{-1}$, corresponding to the limit $C_{pvk} \rightarrow \infty$ of the total device capacitance $\left(\frac{1}{C_{htl}} + \frac{1}{C_{pvk}} + \frac{1}{C_{etl}}\right)^{-1}$. Note that the resistances of the individual layers do not affect the capacitance signal of short-circuited PSCs [32].

When plotting $\omega \frac{\partial C}{\partial \omega}$ vs. $\omega$, the low-frequency increase of capacitance becomes a peak at a characteristic frequency $\omega_0$. If the increase of the capacitance is produced by non-degenerated (for simplicity) defects states in the perovskite bandgap, $\omega_0$ is given by,

$$\omega_0 = N_{C,V}\, c_{n,p} \exp\left(\frac{-E_t}{k_b T}\right), \tag{1}$$

where $E_t$ correspond to the energy position of the defects from the nearest band edge, $N_{C,V}$ is the effective density of states for electrons or holes, and $c_{n,p}$ is the defect capture coefficient for electrons or holes [36]. We observe in the numerical simulation that this analytical equation is valid for PIN devices, such as typical PSCs (see Fig. S2).

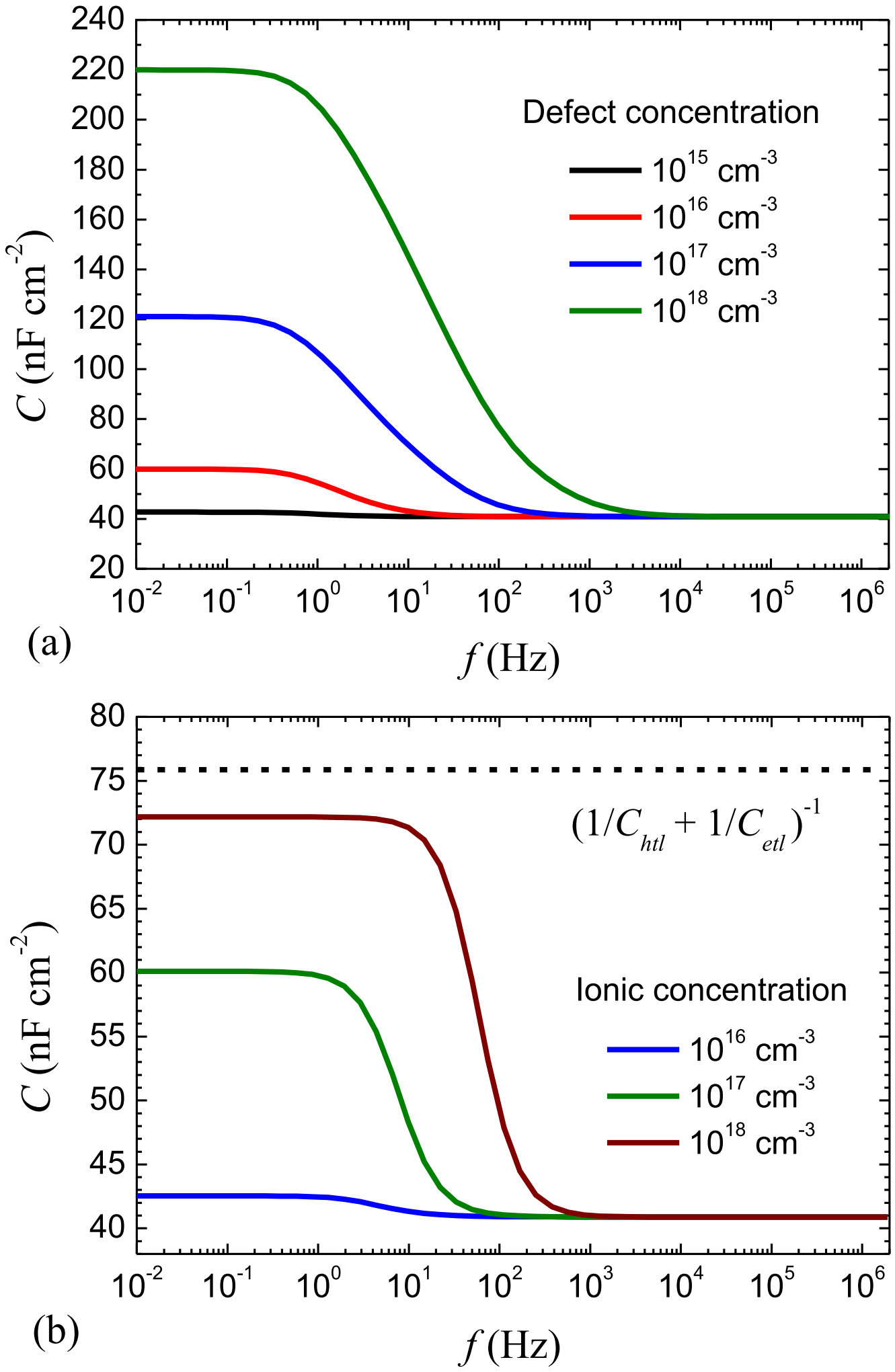

**Fig. 1.** Numerical simulation of the capacitance of a short-circuited PIN solar cell at 300K. The basic parameters of the device were extracted from Table S1 of Ref. [32]. In (a), the low-frequency capacitance is produced by adding to the intrinsic layer different concentrations of donor-like traps at 0.7 eV from the conduction band edge with capture cross sections $\sigma_n^c = \sigma_p^c = 10^{-13}$ cm$^2$. Mid-gap is 0.8 eV from the band edges. In (b), the low-frequency capacitance is produced by adding different concentrations of anions (acceptors) with a mobility of $10^{-8}$ cm$^2$ V$^{-1}$ s$^{-1}$ to the perovskite layer. As the ion concentration increases, the low frequency capacitance approaches the series addition of the geometrical capacitances of the transport layers (dashed line).

If the low-frequency capacitance is produced by a mobile dopant (or ionic species) in the perovskite layer, we could not find a reliable analytical formula for $\omega_0$ in the bibliography. Consequently, we use Silvaco Atlas to perform a drift-diffusion numerical simulation of a semiconductor with mobile dopants/ions to extract approximate formulas. The simulated devices used to derive the formulas are shown in Fig. 2. They consist of one-dimensional (1D) semiconductors with a mobile ionic species and metallic contacts at their extremes. The work functions of each material $\phi = E_{VL}$ - $E_F$ are given before contact, where $E_{VL}$ is the vacuum-level energy, and $E_F$ is the Fermi-level energy of each material. Fig. 2a correspond to a semiconductor with two flat band contacts ($\phi_1^m = \phi_2^m = \phi_s$), i.e. with no intrinsic electric field inside the semiconductor; Fig. 2b correspond to a p-type semiconductor with a flat band contact ($\phi_1^m = \phi_s$) and a rectifying one ($\phi_2^m < \phi_s$). In this case, there is an intrinsic electric field in the bulk of the semiconductor due to the different work functions of the contacts ($\phi_2^m < \phi_1^m$).

We observe that $\omega_0$ does not depend on the semiconductor electron or hole mobilities, dopant concentration, recombination parameters, and charge sign of the mobile ionic species. On the other hand, we observe that it depends on the ionic species concentration $N_s$, mobility $\mu_s$, and charge $qZ_s$. It also depends on the semiconductor thickness $d$, permittivity $\varepsilon = \varepsilon_r \varepsilon_0$, temperature $T$, and electrostatic potential difference between its interfaces $\Delta V_i$. $\Delta V_i$ depends on the work function difference between the contacts and the external bias. In Fig. 2(a), $\Delta V_i$ is produced by applying an external potential in one of the contacts ($\Delta V_i = V_1 - V_2$). In Fig. 2b, $\Delta V_i$ is produced by changing the work function of the second contact ($\Delta V_i = \phi_1^m - \phi_2^m$). As expected, we observe that $\omega_0$ just depends on $\Delta V_i$, independently of how it is produced.

In order to extract a formula for the dependence of $\omega_0$ on these parameters, we changed them one by one from the initial values given in Table 1. The variation ranges of the parameters are also presented in the table. More details of this numerical simulation are given in Section 1.C of the SI.

We extract a formula for $\Delta V_i = 0$ V, where the mobile dopants are initially distributed uniformly (symmetrical ionic distribution), and another formula for $|\Delta V_i| \geq 0.2$ V, where the ionic species are accumulated next to the oppositely charged contact (asymmetrical ionic distribution).

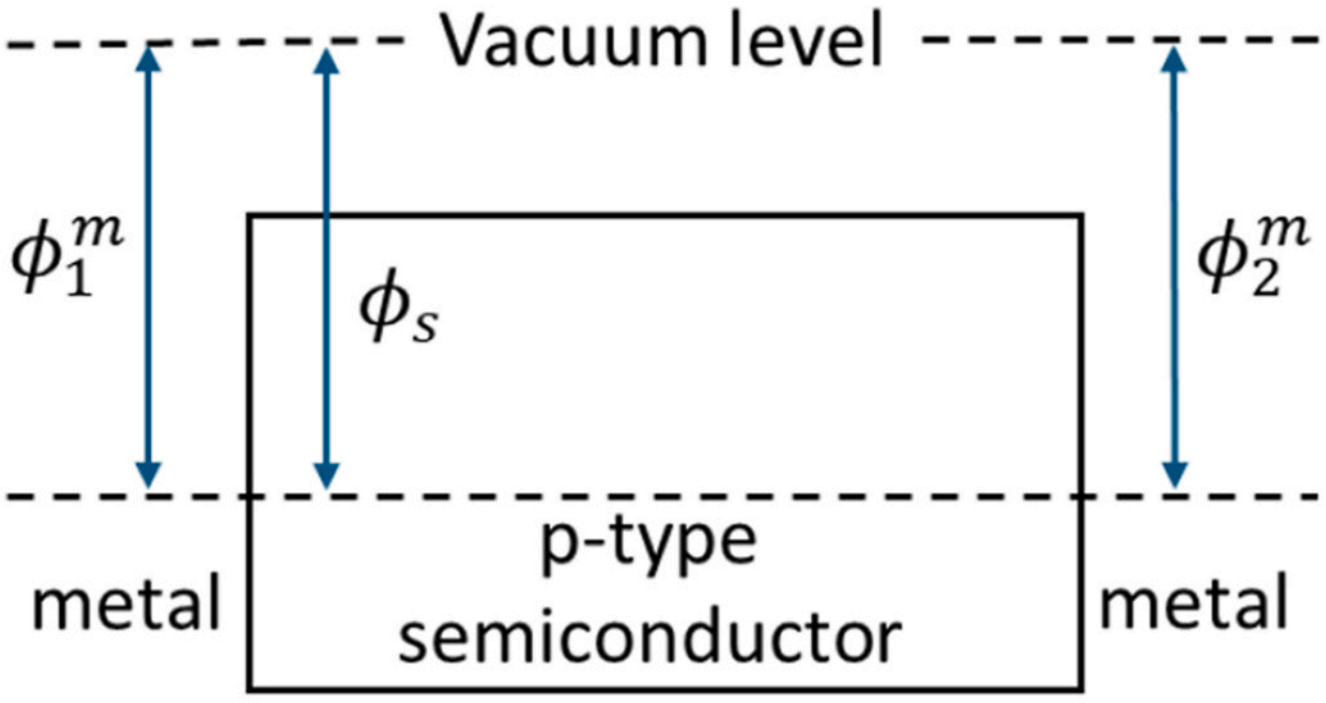

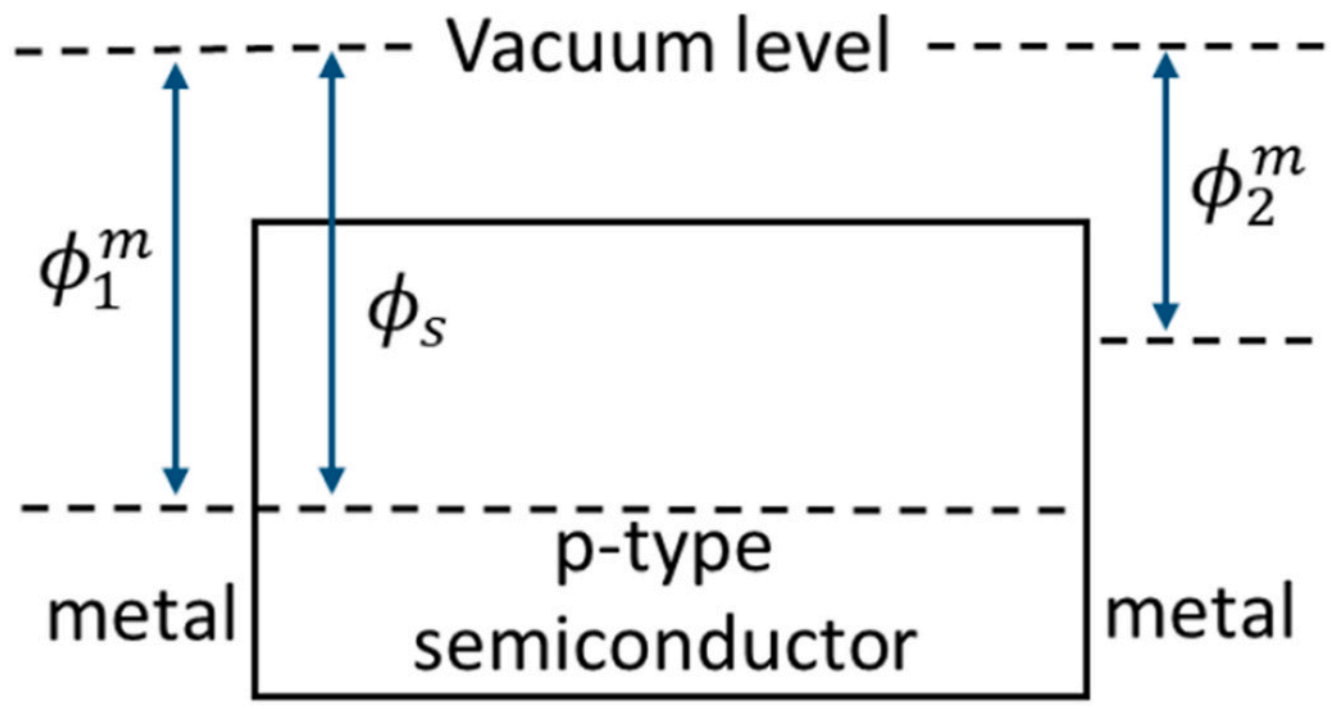

**Fig. 2.** Simulated devices used for extracting formulas (2) and (3). They consist of a 1-D semiconductor contacted in its extremes with a mobile ionic species. The work functions $\phi$ of each material are given before they are contacted. (a) Symmetrical device, contacted with two flat band contacts ($\phi_1^m = \phi_2^m = \phi_s$). There is no intrinsic electric field in the semiconductor. (b) Asymmetrical device, contacted with a flat band contact ($\phi_1^m = \phi_s$) and a rectifying one ($\phi_2^m < \phi_s$). There is an intrinsic electric field in the semiconductor's bulk due to the different work functions of the contacts.

**Table 1**
Initial values and variation ranges of the parameters used in the numerical simulation for deducing Eqs. (2) and (3). We simulate a semiconductor with a completely ionized mobile dopant.

| Parameter [unit] | Symbol | Initial | Range |
|---|---|---|---|
| Species mobility [$cm^2 V^{-1} s^{-1}$] | $\mu_s$ | $10^{-6}$ | $10^{-7}$ - $10^{-5}$ |
| Species concentration [$cm^{-3}$] | $N_s$ | $10^{17}$ | $10^{17}$ - $10^{19}$ |
| Species charge number | $Z_s$ | 1 | 1 - 4 |
| Relative permittivity | $\varepsilon_r$ | 20 | 10 - 80 |
| Temperature [K] | $T$ | 300 | 200 - 400 |
| Layer thickness [μm] | $d$ | 0.5 | 0.25 - 8 |
| Potential difference [V] | $\Delta V_i$ | 0.8 | 0.2 - 1.2 |

For a semiconductor with no potential difference between its interfaces ($\Delta V_i = 0$ V), such as Fig. 2a device under short-circuit condition ($V_{DC} = 0$ V), we obtain

$$\omega_0 = \alpha_1 \frac{q\, Z_s\, \mu_s}{d^{1.15}} \sqrt{\frac{N_s}{\varepsilon (k_b T)}}, \tag{2}$$

where the proportionality constant is $\alpha_1 = 0.014$ V cm$^{0.15}$. For a semiconductor with a potential difference between it interfaces ($|\Delta V_i| \geq 0.2$ V), such as Fig. 2b device under short-circuit condition or Fig. 2a device under an external bias, we obtain

$$\omega_0 = \alpha_2 \frac{\mu_s\, \Delta V_i^{0.55}}{(k_b T)\, d^{1.15}} \sqrt{\frac{(q\, Z_s)^3\, N_s}{\varepsilon}}, \tag{3}$$

where the proportionality constant is $\alpha_2 = 0.012$ cm$^{0.15}$V$^{0.95}$. Both formulas present the same dependence on $\mu_s$, $N_s$, $d$, and $\varepsilon$, but not on $Z_s$ and $T$.

Fig. 3 presents the $\omega_0$ values extracted from the simulation while changing one by one the parameters of Table 1, when (a) $\Delta V_i = 0$ V and (b) $|\Delta V_i| \geq 0.2$ V (symbols). The solid lines in (a) and (b) correspond to the fit of these values with Eqs. (2) and (3) for extracting $\alpha_1$ and $\alpha_2$, respectively.

As usual, we assume an exponential temperature dependence of the ionic species mobility with temperature $\mu_s = \mu_s^0 \exp[-E_\mu / (k_b T)]$ [14,30]. Therefore, Eqs. (2) and (3) become:

$$\omega_0 = \alpha_1 \frac{q\, Z_s\, \mu_s^0}{d^{1.15}} \sqrt{\frac{N_s}{\varepsilon (k_b T)}} \exp[-E_\mu / (k_b T)], \tag{4}$$

$$\omega_0 = \alpha_2 \frac{\mu_s^0\, \Delta V_i^{0.55}}{(k_b T)\, d^{1.15}} \sqrt{\frac{(q\, Z_s)^3\, N_s}{\varepsilon}} \exp[-E_\mu / (k_b T)]\ . \tag{5}$$

According to these results, neither of the two analytical formulas previously used for estimating the defect conductivity [30] and diffusion coefficient [14] from $\omega_0$ is consistent with the standard drift-diffusion modeling of transport for this range of parameters. The mentioned analytical formulas from Refs. [14,30] are obtained by assuming that the drift contribution to the ionic current can be neglected. The presence of the layer thickness and the potential difference between the interfaces in the new formulas indicates the contrary. From now on, we assume that ionic species have a single charge ($Z_s = 1$).

Formulas (4) and (5) were extracted from a one-layer device with mobile ions. In a short-circuited PIN device, such as a PSC, we need to add the geometrical capacitances of the hole and electron transport layers ($C_{htl}$ and $C_{etl}$) to obtain the total capacitance of the device, $C(\omega) = \left(\frac{1}{C_{pvk}(\omega)} + \frac{1}{C_{htl}} + \frac{1}{C_{etl}}\right)^{-1}$. We observe numerically that the relations (4) and (5) are also valid in this case, although the proportionality constants $\alpha_1$ and $\alpha_2$ depend on $C_{htl}$ and $C_{etl}$ values. The deduction of Eqs. (4) and (5) from an approximate analytical model such as the one developed in Ref. [37] is beyond the scope of this paper.

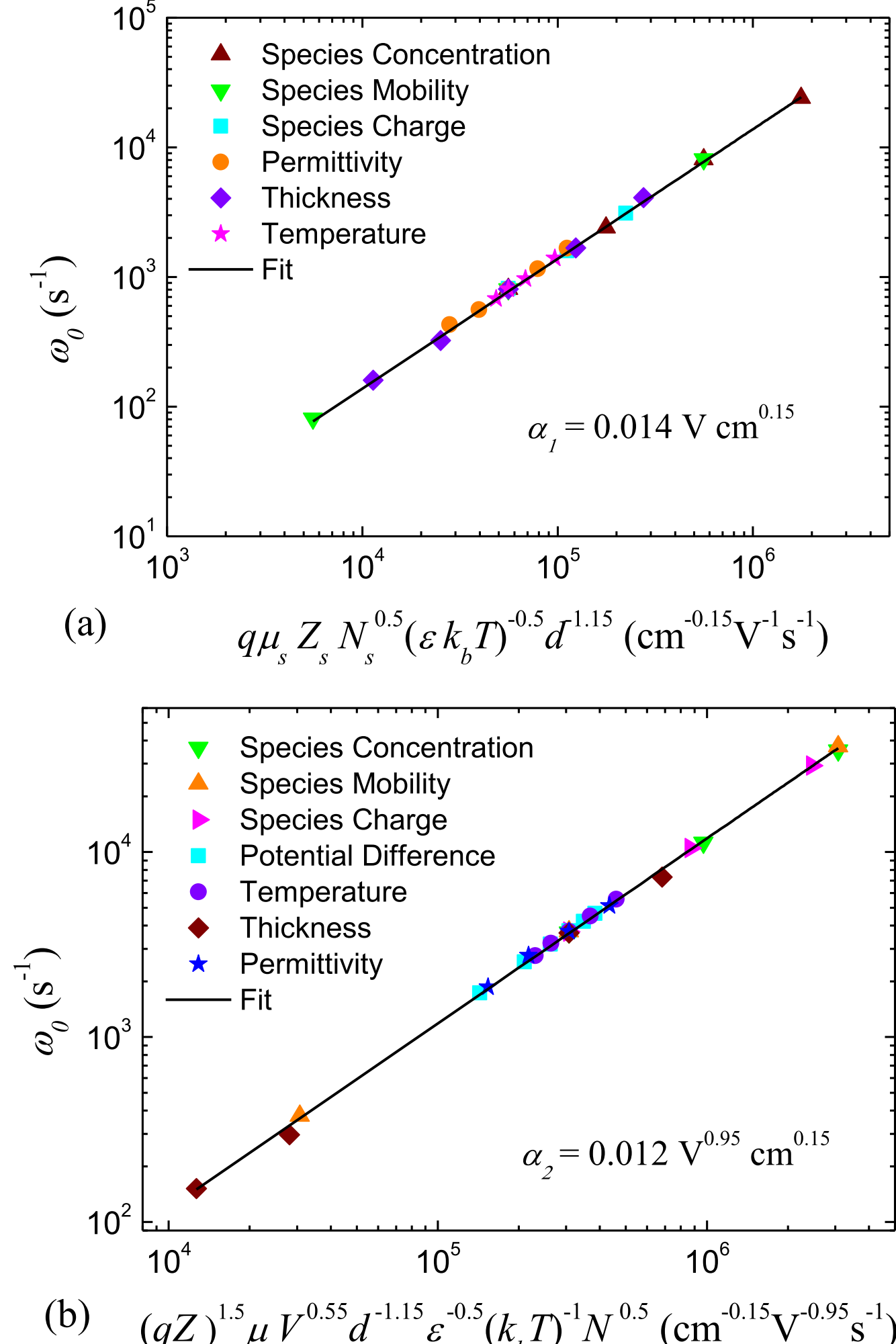

**Fig. 3.** $\omega_0$ as a function of the material parameters displayed in Table 1. $\omega_0$ corresponds to the frequency of the $-\omega \frac{\partial C}{\partial \omega}$ peak. The symbols were extracted from the numerical simulation while changing one by one the parameters, when (a) $\Delta V_i = 0$ V and (b) $|\Delta V_i| \geq 0.2$ V. The solid lines in (a) and (b) correspond to the fit of these values with Eqs. (2) and (3) for extracting $\alpha_1$ and $\alpha_2$, respectively.

### 2.2. Steady-state and quasi-steady-state techniques in halide perovskites

Steady-state techniques such as SSPC and SSPG were originally developed to characterize conventional semiconductors, which lack mobile defects. SSPC and SSPG usually provide information about the majority and minority carrier, respectively. These methods assume a steady-state condition of the material under fixed polarization bias, illumination, and temperature. A general description of the SSPC and SSPG techniques is given in Section 2 of the SI.

Halide perovskites are known for their long transient responses (of several minutes) upon changes in polarization bias, which are typically associated with the migration of defects [38–40]. Similar long transients have also been reported when illumination is suddenly applied in biased devices. They have been interpreted as resulting from the trapping/emission of free charge by deep defects in the bandgap and/or the migration of defects [41,42].

Ionic migration is not included in the standard theory of SSPC and SSPG (see Section 2 of the SI). A recommended method to avoid the effects of defect migration in SSPC and SSPG is to characterize the material under **quasi-steady-state** conditions. In this approach, measurements are performed slowly enough for the electronic distribution to remain in steady state with external excitations, yet fast enough to minimize defect migration [43].

Solar cell devices work under steady-state conditions. Therefore, their Power Conversion Efficiency (PCE) is determined from the **steady-state** current-voltage (IV) measurements under standard conditions (AM1.5, 1 kW m$^{-2}$, 25 °C). Initially, we extract the Open-Circuit voltage ($V_{OC}$), the Short-Circuit current density ($J_{SC}$), and the Fill Factor (FF) from the IV curve. Then, we estimate the PCE using the formula, PCE = ($V_{OC} \times J_{SC} \times$ FF)/$P_{in}$, where $P_{in}$ is the input light power. It is essential that the IV curve corresponds to the steady-state one, at least for voltages below the $V_{OC}$, because the PCE is extracted from this range of voltages.

## 3. Experimental

### 3.1. Samples details

Fig. 4 presents a schematic representation of the solar cell samples: panel (a) shows the top view, while panel (b) displays the cross-sectional view. Each sample consists of 16 independent solar cell devices (labelled SC1 through SC16), as illustrated in Fig. 4a. The layers composing each solar cell device, along with their respective thicknesses, are detailed in Fig. 4b. The optical bandgap of this type of perovskite is in the range 1.65-1.8 eV [10,11]. To minimize environmental stresses [25], all devices were coated with $Al_2O_3$ (30 nm) using atomic layer deposition (Arradiance's GEMStar XT Thermal ALD) prior to characterization. The area of an individual solar cells is 0.08 cm$^2$ but the illuminated area is smaller (0.05 cm$^2$).

The horizontal or lateral samples consist of the same perovskite layer, approximately 550 nm thick, but deposited on regular Corning glass. On top of the perovskite, we evaporate 100 nm of two rectangular gold contacts (see Fig. 5). The samples were also coated with 30 nm of $Al_2O_3$ by atomic layer deposition (using Arradiance's GEMStar XT Thermal ALD) to isolate them from the environment [25]. Fig. 5 shows a schematic of the horizontal samples, with panel (a) depicting the top view and panel (b) showing the cross-sectional view.

### 3.2. Characterization procedure

We characterize three solar cell samples (S2, S3 and S5) and three horizontal devices (H1, H2 and H3). Initially, we perform IV measurement under standard conditions (AM1.5, 1 kW m$^{-1}$, 25 °C) for the 16 solar cells of the **samples S2** and **S5**. Then, we perform TAS on two solar cells (SC1 and SC4) of the **sample S3**.

We apply different characterization techniques to three horizontal devices. In the **H1 device**, we initially measure the dark current transients upon the application of a polarization bias and then upon turning on the (740 nm) LED light under a constant polarization bias. Next, we perform IV measurements at room temperature, scanning the bias in both directions under dark initially and then under the LED illumination. Finally, we perform IV measurements from 300 to 100 K and vice versa in 20 K steps. For each temperature, we measure initially the dark IV and then the IV under the LED light at two light fluxes (first at 10$^{16}$ and then 5 × 10$^{15}$ cm$^{-2}$s$^{-1}$).

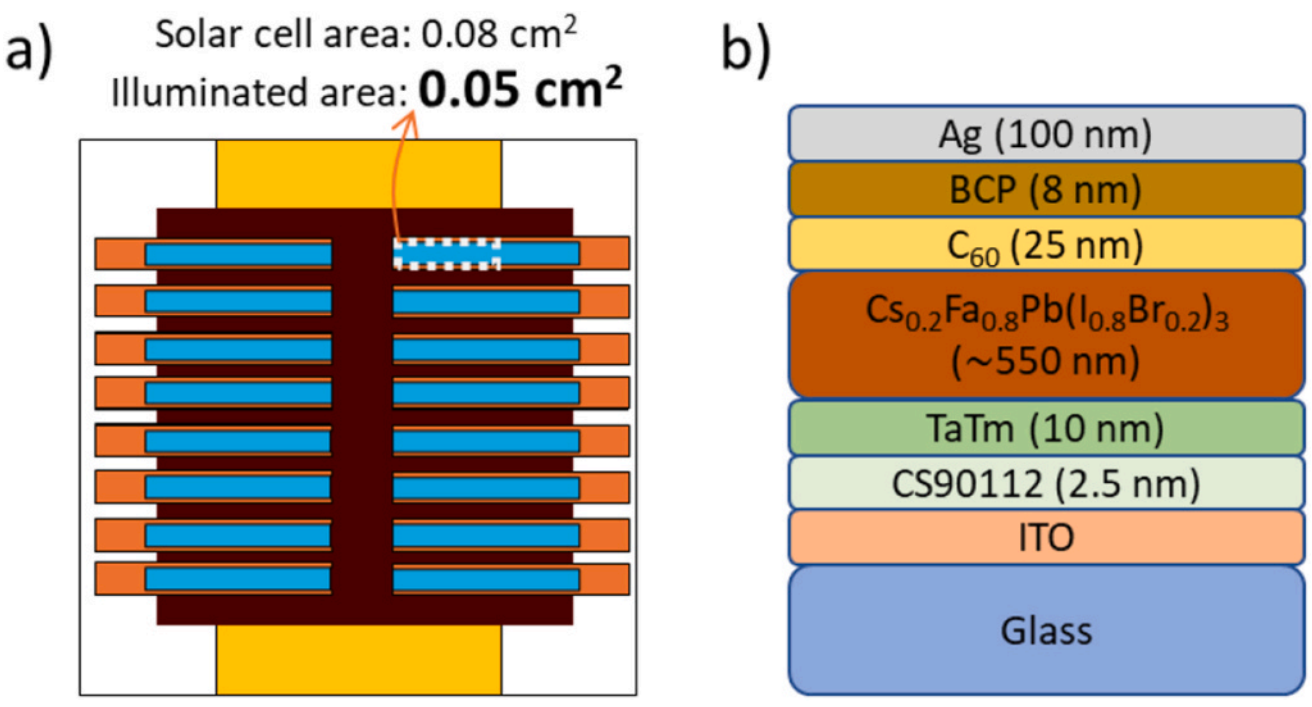

**Fig. 4.** Diagram of the sandwich-type samples, where (a) is the top view and (b) is the cross-sectional view. Each sample consists of sixteen independent solar cell devices, and each solar cell is composed of nine different layers. The top layer ($Al_2O_3$) just isolates the device from the environmental stresses.

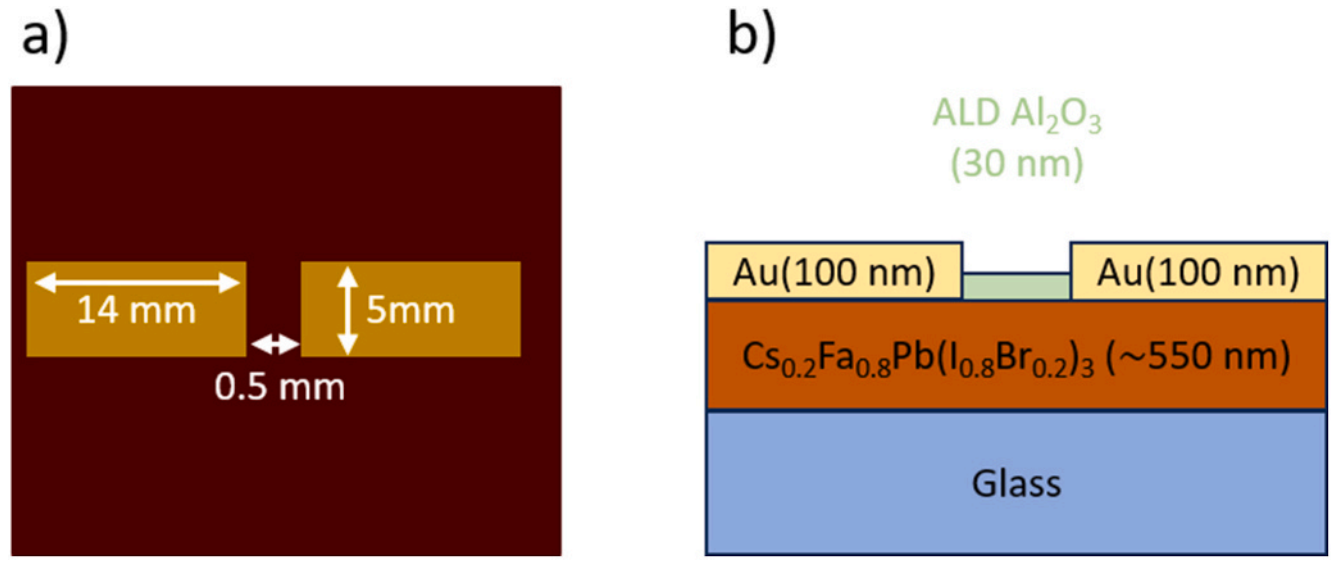

**Fig. 5.** Schematic of the horizontal samples, where (a) is the top view and (b) is the cross-sectional view.

In the **H2 device**, we conducted measurements at various temperatures, following a similar protocol as with the first device. In this case, the bias was scanned in both directions at a significantly slower rate, and dark measurements were entirely separated from those performed under illumination.

In the **H3 device**, we perform SSPG and SSPC measurements at room temperature for different voltages using a 633 nm HeNe laser.

The carrier generation rate under illumination is estimated from the equation

$$G_r = F(1 - R_f)\frac{(1 - e^{-\alpha d})}{d}, \tag{6}$$

where $F$ is the light flux, $R_f$ is the front surface reflectance, and $\alpha$ is the absorption coefficient of the material. From the absorbance of Ref. [10], we estimate $\alpha$ (633 nm) = 4.6 × 10$^4$ cm$^{-1}$ and $\alpha$ (740 nm) = 2.3 × 10$^4$ cm$^{-1}$ for our perovskite. We assume $R_f$ = 0.15 for both wavelengths.

All horizontal and solar cell samples were deposited at ICMol (Spain). The standard IV of the solar cell devices were measured at ICMol and the rest of the characterization measurements were performed at GeePs (France). The procedure followed for the deposition of the solar cells and the complete characterization details are given in Section 3 of the SI.

## 4. Experimental results and discussion

### 4.1. Solar cells

a. Current-voltage (IV) measurements under standard conditions

Fig. 6a and b presents the current density versus voltage (*J-V*) under standard conditions (AM1.5, 1 kW m$^{-1}$, ~25 °C) obtained for samples S2 and S5, respectively. Each plot displays the 16 individual solar cell measurements of each sample represented by differently colored lines. For ensuring accurate PCE determination, *J-V* measurements must be performed slowly enough to capture steady-state currents below $V_{OC}$. Note that the *J-V* curves exhibit negligible hysteresis below the open-circuit voltage. Sample S5 displays more consistent behavior across cells, while sample S2 shows greater variability in the results.

Fig. 7 presents the statistics of the parameters extracted from the *J-V* curves: the Open-Circuit voltage ($V_{OC}$), the Short-Circuit current density ($J_{SC}$), the Fill Factor (FF), and the Power Conversion Efficiency (PCE). Our champion solar cells achieve a PCE of 18% (see Fig. 7c), equal to those reported in Ref. [10] for similar devices. Although our $J_{SC}$ is slightly lower and our $V_{OC}$ is slightly higher than in Ref. [10], those differences likely stem from different transport and passivation layers.

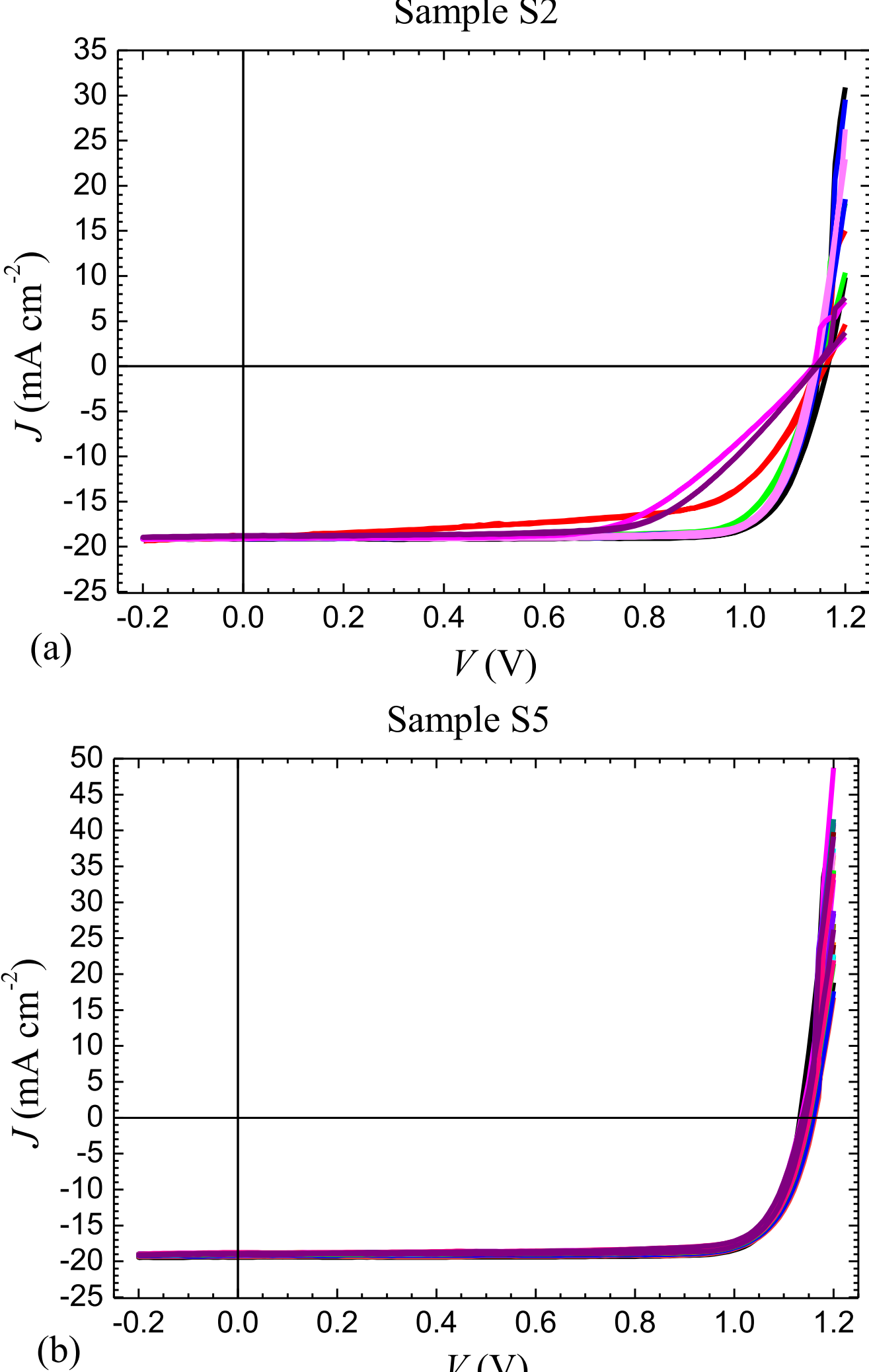

**Fig. 6.** Current density versus voltage under standard conditions (AM1.5, 1 kW m$^{-1}$, ~25 °C) for samples S2 and S5. The different color lines in each plot correspond to the 16 solar cells of each sample. More consistent values were obtained for sample S5. No significant hysteresis is detected in either sample.

The measured $V_{OC}$ (~1.15 V) and $J_{SC}$ (~19 mA cm$^{-2}$) are consistent with the perovskite bandgap of ~1.7 eV (see Section 5), and are typical for this type of wide-bandgap perovskite devices with unoptimized light-management scheme [11].

The dispersion in the *J-V* curves of sample S2 arises mainly from variations in the FF, which ranges from 60% to 81%. This reduction in FF, without significant changes in $V_{oc}$ or $J_{sc}$, suggests slightly higher series resistances and/or lower shunt resistances [44]. This effect is unlikely caused by increased recombination, as that would also induce variations in $V_{OC}$. It is important to note that the sole difference between the two samples is their placement in the substrate holder during deposition.

b. Thermal admittance spectroscopy (TAS)

Fig. 8a and b presents the capacitance measured at different temperatures and frequencies for two short-circuited ($V_{DC} = 0$ V) solar cells of sample S3. Initially, we perform measurements in one solar cell (SC1) from 300 to 200 K and vice versa, in 10 K steps (Fig. 8a). Then, we perform measurements on another cell (SC4) in 5 K steps from 300 to 330 K, and from there to 285 K (Fig. 8b). The slight difference between the measurements performed while increasing (symbols) and decreasing (lines) the temperature could be attributed mainly to thermal inertia. Note that these differences are lower for the second solar cell (SC4), where we use the lower temperature step.

The high-frequency plateaus in Fig. 8a and b correspond to the geometrical capacitance of the device, given by the series addition of the geometrical capacitances of each layer, $C_g = \left(\frac{1}{C_{htl}} + \frac{1}{C_{pvk}} + \frac{1}{C_{etl}}\right)^{-1}$ [30, 32]. We expect that the main contribution to the geometrical capacitance comes from the perovskite layer, due to its higher thickness. Discarding $C_{htl}^{-1}$ and $C_{etl}^{-1}$ in the previous formula, we obtain the inequality $C_{pvk} > C_g$. As expected, the high frequency plateau has practically the same value in both solar cells, $C_g$ (300 K, 10 kHz) = 35 nF cm$^{-2}$. Replacing this value in the previous inequality, $C_{pvk} = \frac{\varepsilon A}{d} > C_g$, we estimate a lower boundary for the perovskite relative permittivity, $\varepsilon_r(300K) > 22$ [30].

A slight increase in the capacitance plateau with temperature is observed, consistent with prior measurements [30]. The high-frequency capacitance drop is attributed to the combined effects of the device series resistance and the characterization system inductance [30]. In the first solar cell (SC1), this drop shifts to lower frequencies, indicating a slightly higher series resistance. This observation correlates with the series resistance variations as the responsible for the dispersion in the IV measurements discussed in the previous section.

Fig. 8c and d presents the derivatives $-dC/d(\ln(\omega))$ vs. $\omega$ of the temperature-decreasing data presented in Fig. 8a and b, respectively. In both solar cells, we observe a low-frequency peak whose frequency position, $\omega_0$, increases with temperature. We could estimate only the values of $\omega_0$ at the three highest temperatures for the first solar cell (see Fig. 8c); values at lower temperatures are below the measurable frequency range of our experimental setup ($f \geq 20$ Hz). The high-frequency peaks are due to the series resistance of the device and the inductance of the characterization system. Note that their positions are practically temperature-independent.

The low-frequency $\omega_0$ values are presented in Fig. 9 as a function of temperature. The solid lines in Fig. 9a and b correspond to the fit of $\omega_0$ vs. $T$ values with Eqs. (1) and (5), respectively. The vertical axis in Fig. 9a is $\omega_0/T^{3/2}$, because we assume $N_{C,V} \propto T^{3/2}$ and that $c_{n,p}$ is temperature independent. The vertical axis in Fig. 9b is $\omega_0 T$, because we assume that $\mu_s^0$, $\Delta V_i$, $d$, $N_s$ and $\varepsilon$ are temperature independent.

In both cases (Fig. 9a and b), we obtain an excellent fit of the experimental data. Therefore, we cannot distinguish in this way between the processes that originate the capacitance signal (free-charge trapping/emission or migration by/of defects). Note that due to the different temperature dependences in the vertical axes of the figures, we obtain slightly different activation energies from the same experimental data.

From the slopes of the fits in Fig. 9a, we would extract defect energy levels of 0.33 and 0.21 eV relative to the nearest band for the perovskites of the first and second solar cells, respectively. Assuming the effective free electron or holes concentration $N_{C,V}(300\ K) = 2 \times 10^{18}$ cm$^{-3}$ [30, 32], we would obtain from the intercept that $c_{n,p}$ is $2 \times 10^{-10}$ and $9 \times 10^{-13}$ cm$^3$s$^{-1}$ for the perovskite defects of the first and second device, respectively. Using these parameters, we attempt to reproduce the experimental capacitance shown in Fig. 8a. Fig. 10 shows the simulated capacitance-frequency response of a short-circuited PSC at 300 K, incorporating (acceptor-like) defects at 0.33 eV from the conduction band (with capture coefficients $c_n = c_p = 2 \times 10^{-10}$ cm$^3$s$^{-1}$). In this case, we slightly adjusted the parameter values of Table S1 of Ref. [32] to match our solar cell configuration (Fig. 4b). We increased the perovskite bandgap and layer thickness, and we changed the permittivity of each layer to reproduce the high-frequency capacitance plateau. We observed that the capacitance signal does not depend on the perovskite bandgap. The final parameter values are provided in the figure caption. Fig. 10a presents the capacitance spectra for $10^{18}$ cm$^{-3}$ defects as we increase the doping of the transport layers. Fig. 10b presents the capacitance spectra for the maximum doping of the transport layers ($10^{17}$ cm$^{-3}$) as we increase the defect concentration. The low-frequency capacitance step increases with the transport layers doping and defect concentration. The simulated capacitance resembles the experimental data (compare Fig. 10 with the 300 K isotherms of Fig. 8a), although the low-frequency capacitance step is significantly lower in the simulation, even with extremely high defect concentrations such as $10^{20}$ cm$^{-3}$ (see Fig. 10b). The low-frequency capacitance step is even lower if we add the defects at 0.33 eV from the valence band, use donor-like defect states, or use the activation energy of the second solar cell (0.21 eV). Therefore, we conclude that it is not possible to produce such high value of the low-frequency capacitance with such low defect activation energy (0.33 or 0.21 eV). High values of the low-frequency capacitance, such as those presented in Fig. 1a, are obtained with defect activation energies higher than 0.6 eV (as it is shown in Fig. S1).

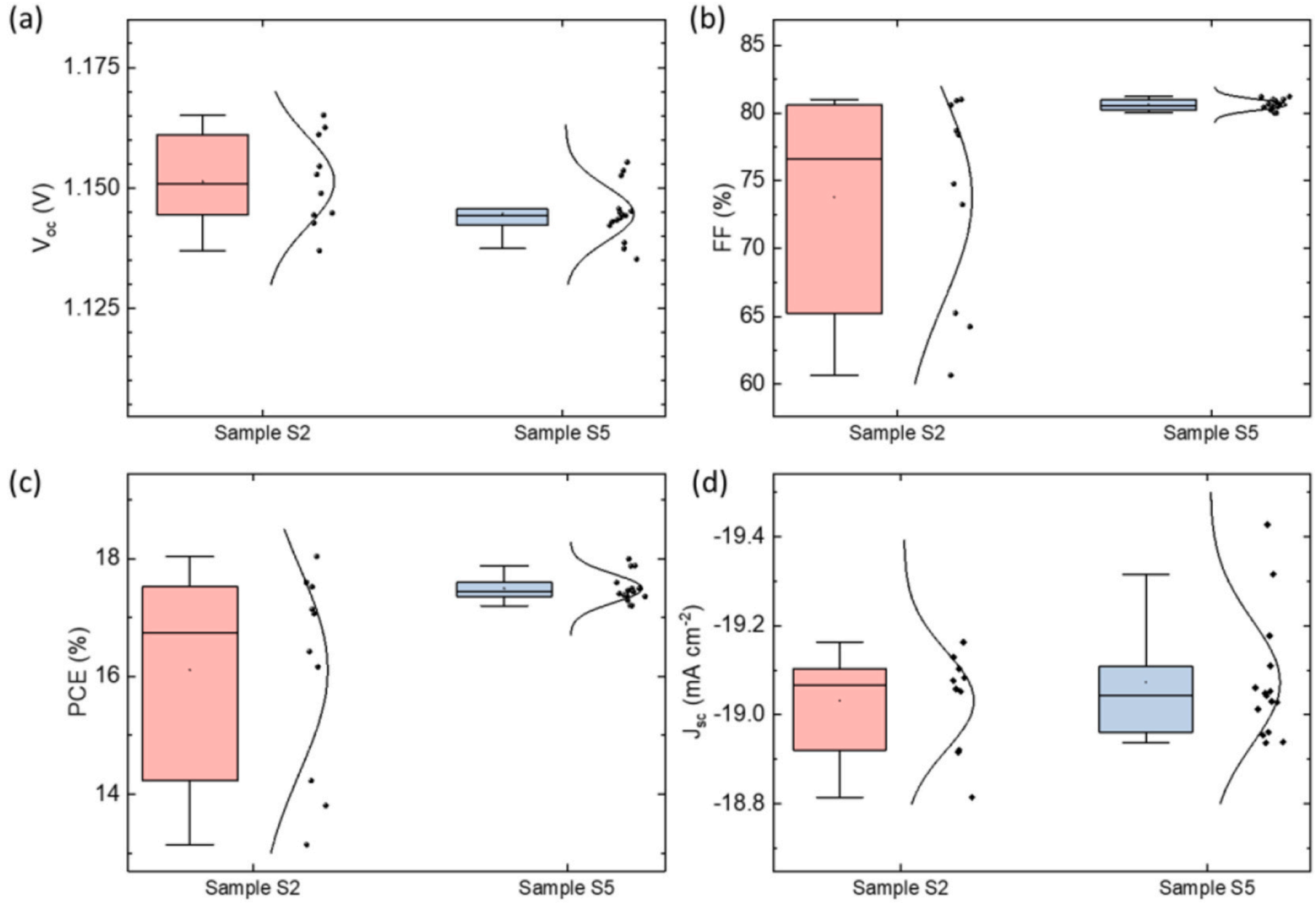

**Fig. 7.** Statistics of JV parameters for samples S2 and S5: (a) Open Circuit voltage ($V_{OC}$), (b) Field Factor (FF), (c) Power Conversion Efficiency (PCE) and (d) Short-Circuit current density ($J_{SC}$). The source of spreading in sample S2 results comes mainly from the FF.

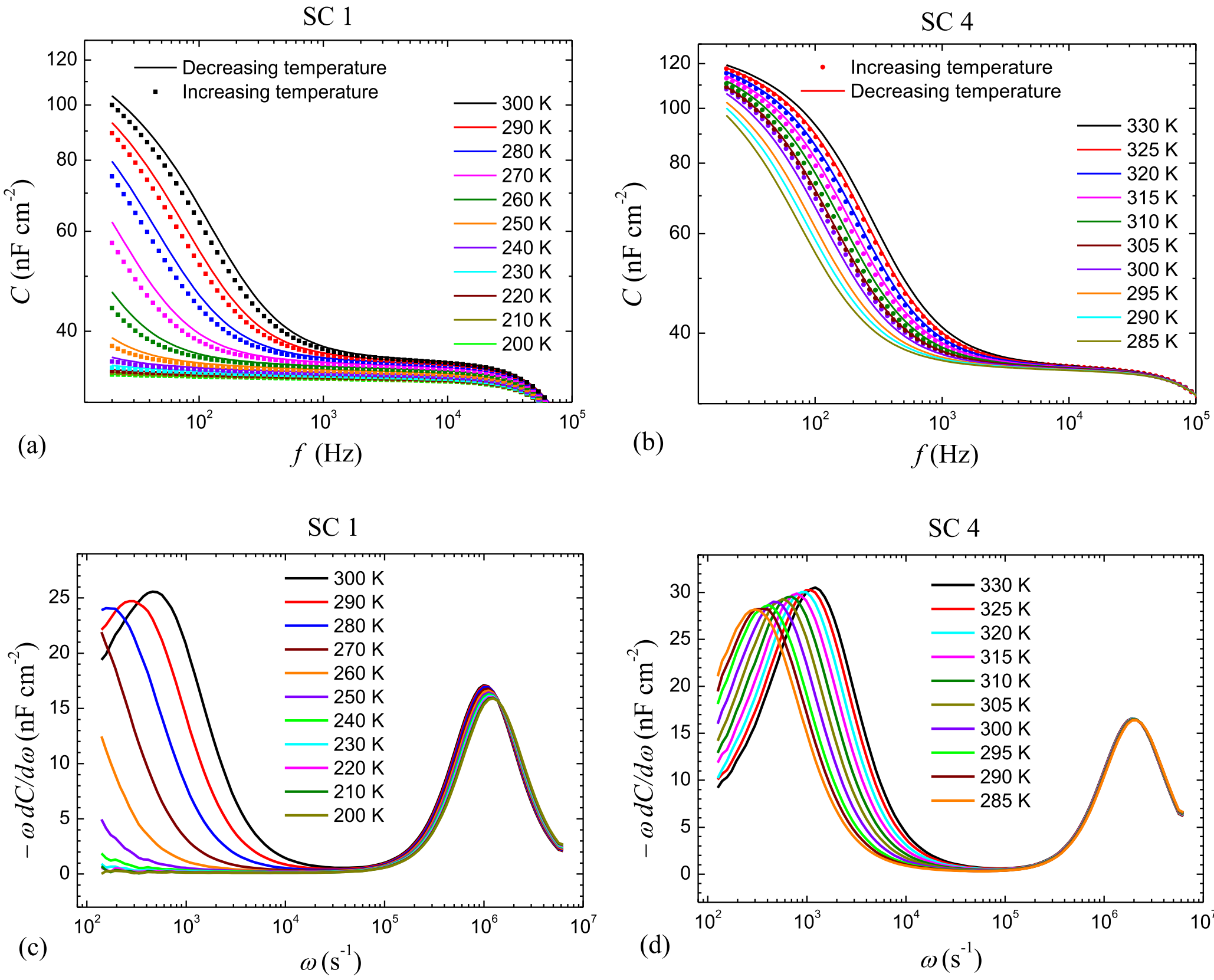

**Fig. 8.** Capacitance vs. frequency at different temperatures of two short-circuited ($V_{DC} = 0$ V) solar cells of sample S3, (a) SC1 and (b) SC4. The slight difference between the measurements performed while increasing (symbols) and decreasing (lines) the temperature could be attributed mainly to thermal inertia. Negative derivative of the temperature decreasing capacitance with respect to the frequency logarithm of (c) SC1 and (d) SC4, plotted against the angular frequency ($\omega = 2\pi f$). The high-frequency peaks do not depend on temperature. The low-frequency peaks shift to higher frequencies as temperature increases; these peak frequencies are plotted against temperature in Fig. 9 for extracting their activation energies.

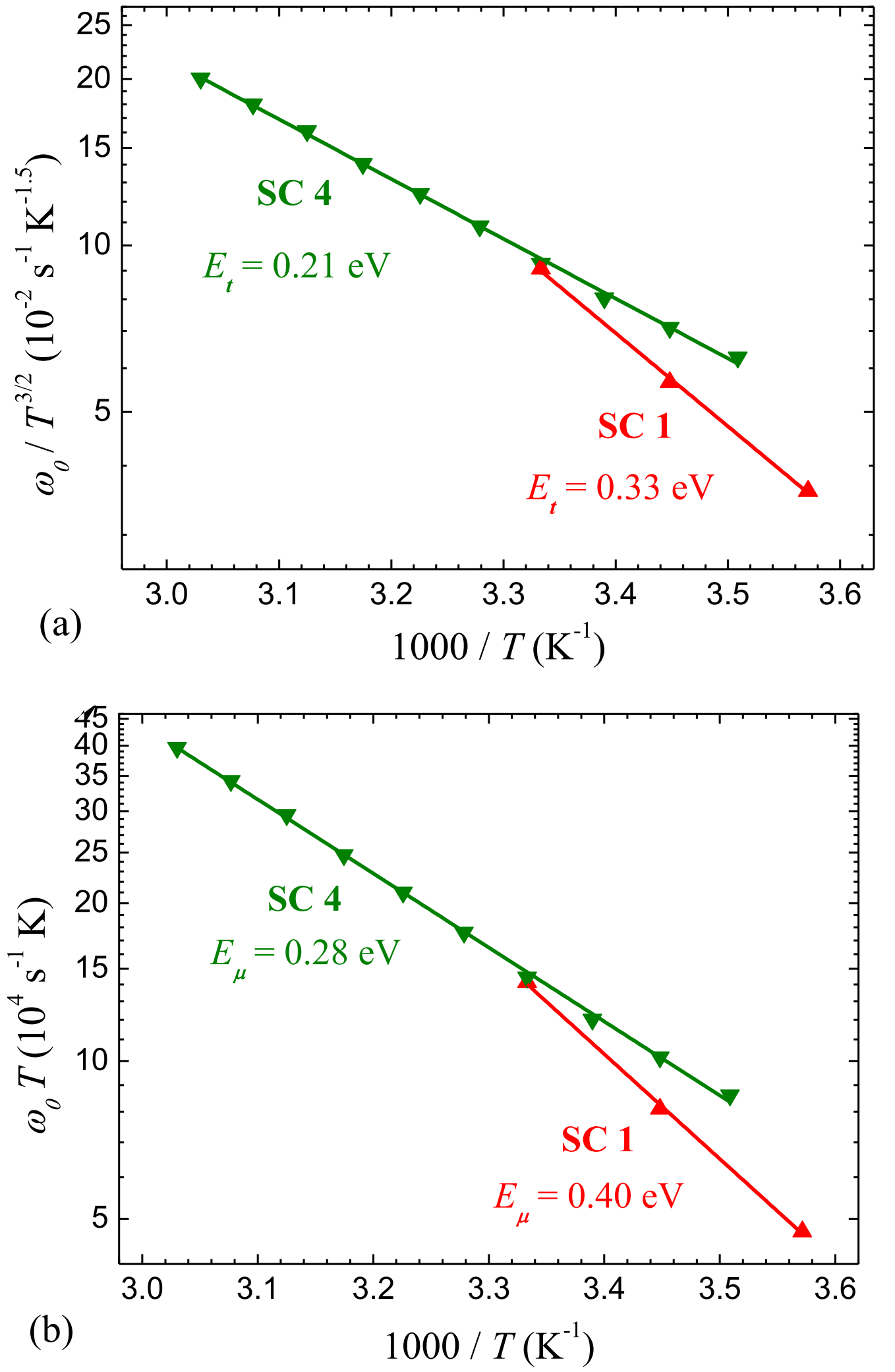

**Fig. 9.** Arrhenius plots of $\omega_0$, representing the angular frequency extracted from the experimental low-frequency peaks in Fig. 8c and d. Upward (red) and downward (green) triangles correspond to two different solar cells of the same sample (SC1 and SC4, respectively). Panels (a) and (b), show the experimental data fitted with Eqs. (1) and (5), respectively. The temperature dependence of the vertical axis follows the respective equation. Both fits match the experimental data excellently, though with slightly different activation energies.

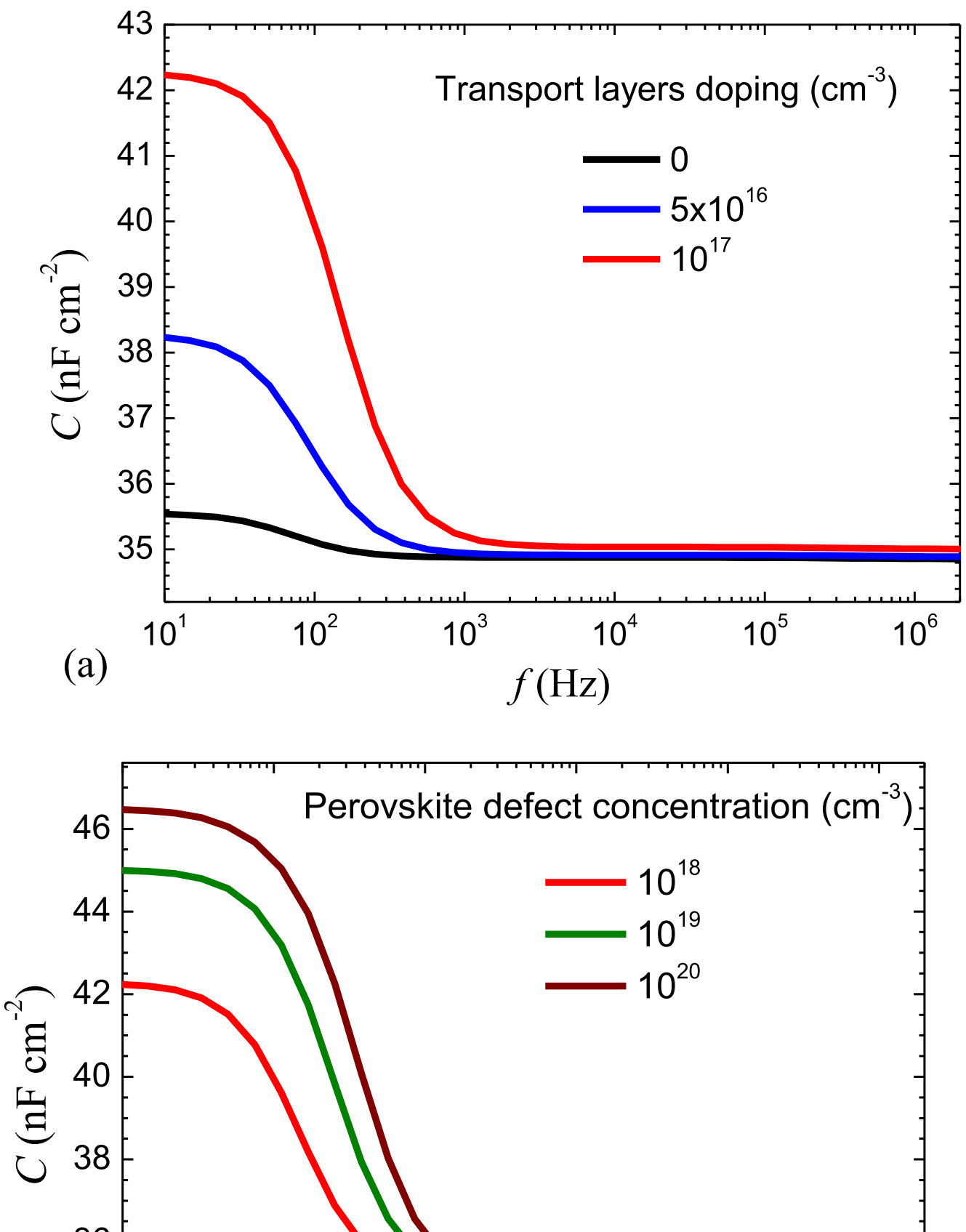

**Fig. 10.** Simulated capacitance vs. frequency of a short-circuited perovskite solar cell at 300 K with acceptor-like defects at 0.33 eV from the conduction band with capture coefficients $c_n = c_p = 2 \times 10^{-10}$ cm$^3$s$^{-1}$. The remaining device parameters correspond to those of Table S1 of Ref. [32], with the exception of the perovskite thickness (550 nm), bandgap (1.65 eV), relative permittivity (33), and permittivities of the transport layers (4). In (a), the defect concentration is $10^{18}$ cm$^{-3}$, and in (b), the transport layers' doping is $10^{17}$ cm$^{-3}$.

From the slopes of the fits in Fig. 9b, we extract ionic mobility activation energies of 0.40 and 0.28 eV for the perovskites of the first and second solar cells, respectively. Neglecting the variation of $\alpha_2$ due to the capacitance of the transport layers and assuming typical values of $\Delta V_i = 0.80$ V and $\varepsilon_r = 33$ for the perovskite layer [30,32,45], we extract from the intercept that $\sqrt{N_s}\,\mu_s$(300 K) is equal to 40 and 80 cm$^{0.5}$V$^{-1}$s$^{-1}$ in the first and second device perovskite, respectively. We observe in Fig. 1b that to produce a significant increase in the low-frequency capacitance, we need to add at least $10^{16}$ cm$^{-3}$ mobile dopants. According to Refs. [46,47], stable devices with high efficiencies never reach ionic concentrations of $10^{19}$ cm$^{-3}$. From these two extreme values for $N_s$ and our estimations of $\sqrt{N_s}\,\mu_s$(300 K) from Eq. (5), we could estimate that $\mu_s$(300 K) should be between $10^{-8}$ and $8 \times 10^{-7}$ cm$^2$ V$^{-1}$ s$^{-1}$ for this perovskite.

A more accurate estimation of $N_s$ and $\mu_s$ should be achieved by reproducing the experimental capacitance-versus-frequency results through numerical simulations. In this approach, we extract information not only from $\omega_0$ vs. $T$ but also from the capacitance values. We focus on simulating the highest temperature isotherm (330 K) presented in Fig. 8b (corresponding to SC4), as it reveals more details in the low-frequency region. However, even at this temperature, the low-frequency capacitance plateau falls below the measurable frequency range of our experimental setup. We extrapolate from the lowest frequency data that the plateau should be slightly above 126 cm$^{-2}$ nF. This value cannot be obtained using the permittivity values reported in Table S1 of Ref. [32], since the low-frequency capacitance saturates to $\left(\frac{1}{C_{HTL}}+\frac{1}{C_{ETL}}\right)^{-1}=76\ \mathrm{cm^{-2}}$ nF in this case, as shown in Fig. 1b. To resolve this discrepancy, we adjusted the permittivity values, increasing those of the transport layers and decreasing that of the perovskite, aiming to fit both high- and low-frequency experimental capacitance values simultaneously. Note that in this case, the doping of the transport layers does not have any influence in the low-frequency capacitance.

Fig. 11a compares the experimental 330 K isotherm from Fig. 8b (symbols) with two different fits obtained from the numerical simulation. The thermal activation energy of the ionic mobility is 0.28 eV and was extracted from Fig. 9b. The solid red line is obtained with a PSC containing $2\times10^{17}\ \mathrm{cm^{-3}}$ ions in the perovskite layer with a mobility of $1.1\times10^{-7}\ \mathrm{cm^2\,V^{-1}\,s^{-1}}$ at 300 K and relative permittivity of 27 for the perovskite layer and 7 for the transport layers. In contrast, the dashed blue line corresponds to a PSC containing $2\times10^{18}\ \mathrm{cm^{-3}}$ ions in the perovskite layer with a mobility of $1.2\times10^{-8}\ \mathrm{cm^2\,V^{-1}\,s^{-1}}$ at 300 K and relative permittivity of 29 for the perovskite layer, 6 and 5 for the HTL and ETL, respectively. Note that by changing the permittivities of the layers, we can produce the same capacitance step with two different ionic concentrations. As we explain in Section 2.1, the proportionality constants $\alpha_1$ and $\alpha_2$ of Eqs. (4) and (5) depend on the thickness and permittivities of the transport layers.

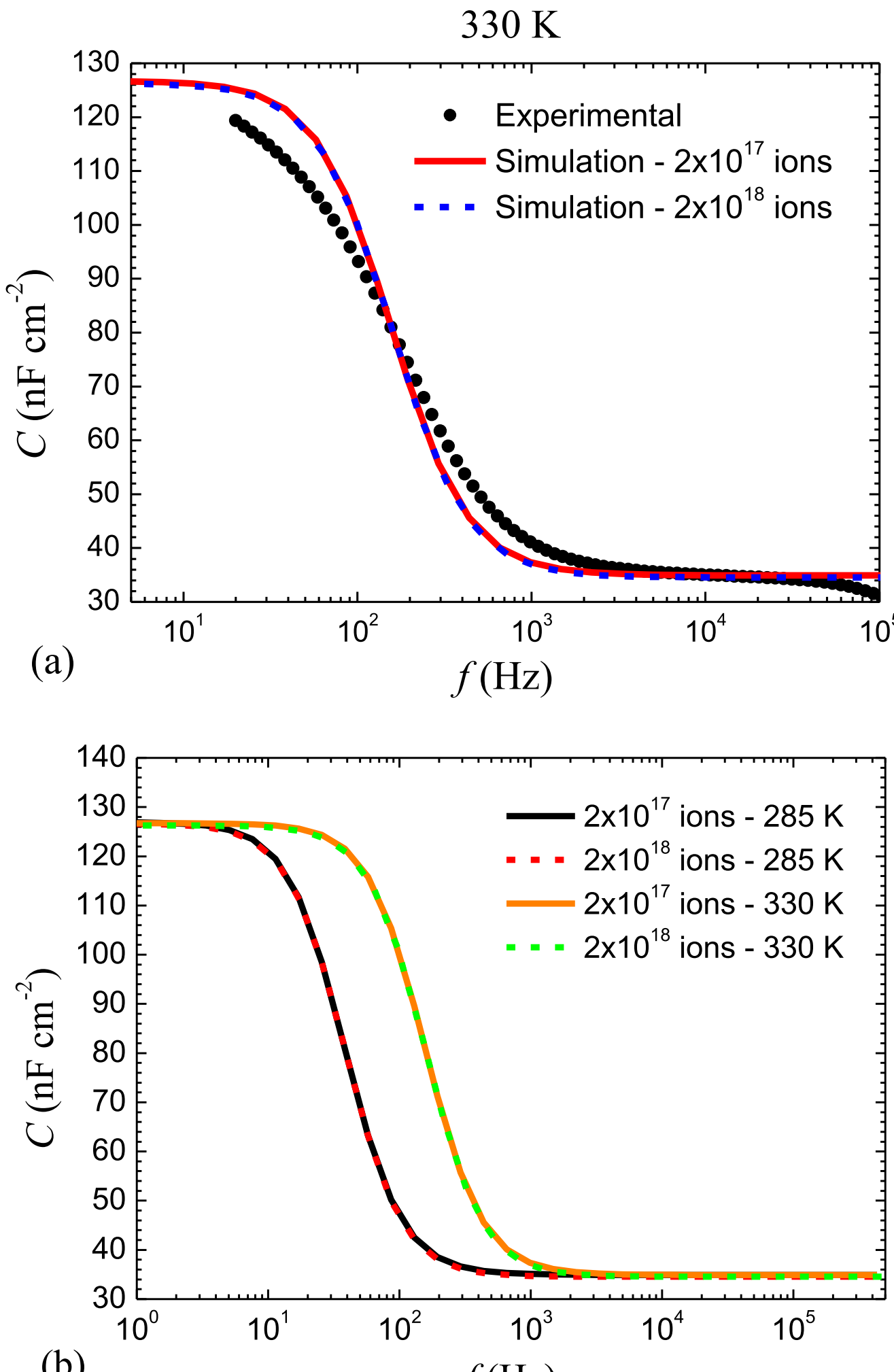

**Fig. 11.** (a) Experimental (SC4) and simulated capacitance vs. frequency of a short-circuited PSC at 330 K. The thermal activation energy of the ionic mobility is 0.28 eV and was extracted from Fig. 9b. The solid red line is obtained with a PSC containing $2\times10^{17}\ \mathrm{cm^{-3}}$ ions in the perovskite layer with a mobility of $1.1\times10^{-7}\ \mathrm{cm^2\,V^{-1}\,s^{-1}}$ at 300 K and relative permittivity of 27 for the perovskite layer and 7 for the transport layers. The dashed blue line is obtained with a PSC containing $2\times10^{18}\ \mathrm{cm^{-3}}$ ions in the perovskite layer with a mobility of $1.2\times10^{-8}\ \mathrm{cm^2\,V^{-1}\,s^{-1}}$ at 300 K and relative permittivity of 29 for the perovskite, 5 and 6 for the electron and hole transport layers, respectively. The rest of the device parameters are those detailed in the caption of Fig. 10. (b) Simulated capacitance obtained with the same parameters at the two extreme experimental temperatures (285 and 330 K).

Fig. 11b presents the simulated capacitance steps at the extreme experimental temperatures (285 and 330 K). The overlap of both curves persists across temperatures, confirming that transport layers do not affect the thermal activation energy of the ionic mobility.

This model successfully reproduces the low-frequency capacitance plateau using reasonable concentrations of mobile ionic species, supporting the conclusion that the observed step arises from mobile ions. A similar conclusion was reported by Futscher et al. [31,34], who combined steady-state and transient capacitance measurements to show that capacitance steps in PSCs typically originate from the movement or oscillation of ionic species.

The numerically simulated curves exhibit a steeper slope than the experimental data (see Fig. 11a); this difference stems from the assumption that all ions possess identical mobility. Fig. S3 presents the experimental 330 K isotherm with two fits obtained with $2\times10^{17}\ \mathrm{cm^{-3}}$ mobile ions. The red line is obtained with all the ions having the same mobility ($1.1\times10^{-7}\ \mathrm{cm^2\,V^{-1}\,s^{-1}}$ at 300 K), and the green curve is obtained with 75% of the ions having a lower mobility ($1.5\times10^{-8}\ \mathrm{cm^2\,V^{-1}\,s^{-1}}$ at 300 K) and 25% having a higher mobility ($4.6\times10^{-7}\ \mathrm{cm^2\,V^{-1}\,s^{-1}}$ at 300 K). The last case fits the experimental data better with a similar average mobility ($1.3\times10^{-7}\ \mathrm{cm^2\,V^{-1}\,s^{-1}}$ at 300 K). A more realistic model would incorporate a Gaussian distribution of ionic mobilities [48,49]. The problem of estimating a unique mobility distribution from the experimental capacitance step is addressed in Ref. [49].

According to Fig. 9b, the temperature-independent pre-exponential factor ($\mu_s^0\sqrt{N_s}$) increases with the activation energy ($E_\mu$). Fitting the pair of $\mu_s^0\sqrt{N_s}$ and $E_\mu$ values with the Meyer-Neldel formula [36], we obtain a Meyer-Neldel energy $E_{MN}$ of 28 meV. The Meyer-Neldel rule has already been observed in TAS measurements of $MAPbI_3$ perovskites by Reichert et al. [14]. After dozens of measurements in samples of different qualities, they report three different $E_{MN}$ values: 28, 30 and 35 meV, associated to cation vacancies $V_A^-$, halide interstitials $X_i^-$, and cation interstitials $A_i^+$, respectively [14]. Although they used a different formula for $\omega_0$ [14], it exhibits the same temperature dependence as Eq. (5), allowing for quantitative comparisons. Despite being different perovskites, our $E_{MN}$ experimental value of 28 meV matches their lowest reported value [14].

Two possible origins have been proposed to explain the validity of the Meyer-Neldel rule in capacitance measurements of halide perovskites: a distribution of migration pathways, or the multi-phonon emission process that characterizes the hopping of ions [14]. The expression for $\omega_0$ used in Ref. [14] does not depend on ionic concentration, implying that variations in the pre-exponential factor are interpreted solely as changes in the ionic diffusion coefficient (or mobility). However, according to equation (5), such variations may also arise from changes in ionic concentration, adding further complexity to the interpretations of the Meyer–Neldel rule in these measurements.

### 4.2. Lateral devices with coplanar contacts

a. Transient currents

To estimate the appropriate waiting times to reach the quasi-steady state in the horizontal samples, we measure the current transients upon changes in the polarization bias and illumination in sample H1. Fig. 12a corresponds to the current transient under dark conditions upon applying a 10 V polarization bias at $t=0$ min. In Fig. 12b, the sample is initially under dark and 10 V polarization. At about $t=10.5$ min, we turn on the (740 nm) LED with $10^{16}\ \mathrm{cm^{-2}s^{-1}}$ photons and measure the current transient for more than 1 h. In both cases, we observe an initial fast current-increasing transient of the order of seconds, followed by a much slower transient of several minutes in the first case (Fig. 12a) and of several tens of minutes in the second case (Fig. 12b). The insets of Fig. 12 correspond to a zoom of the initial part of the transients, where we can observe the fast transient. We associate the fast transient with electronic processes and the slow one with the migration of charged defects. According to the figures, waiting two or 3 s after changing the polarization bias or the light flux should be sufficient to achieve the quasi-steady state condition. Note that waiting for more time could be a source of error, because the migration of ions could induce changes in the bulk's doping, the screening of the applied electric field, etc. [38,39,42].

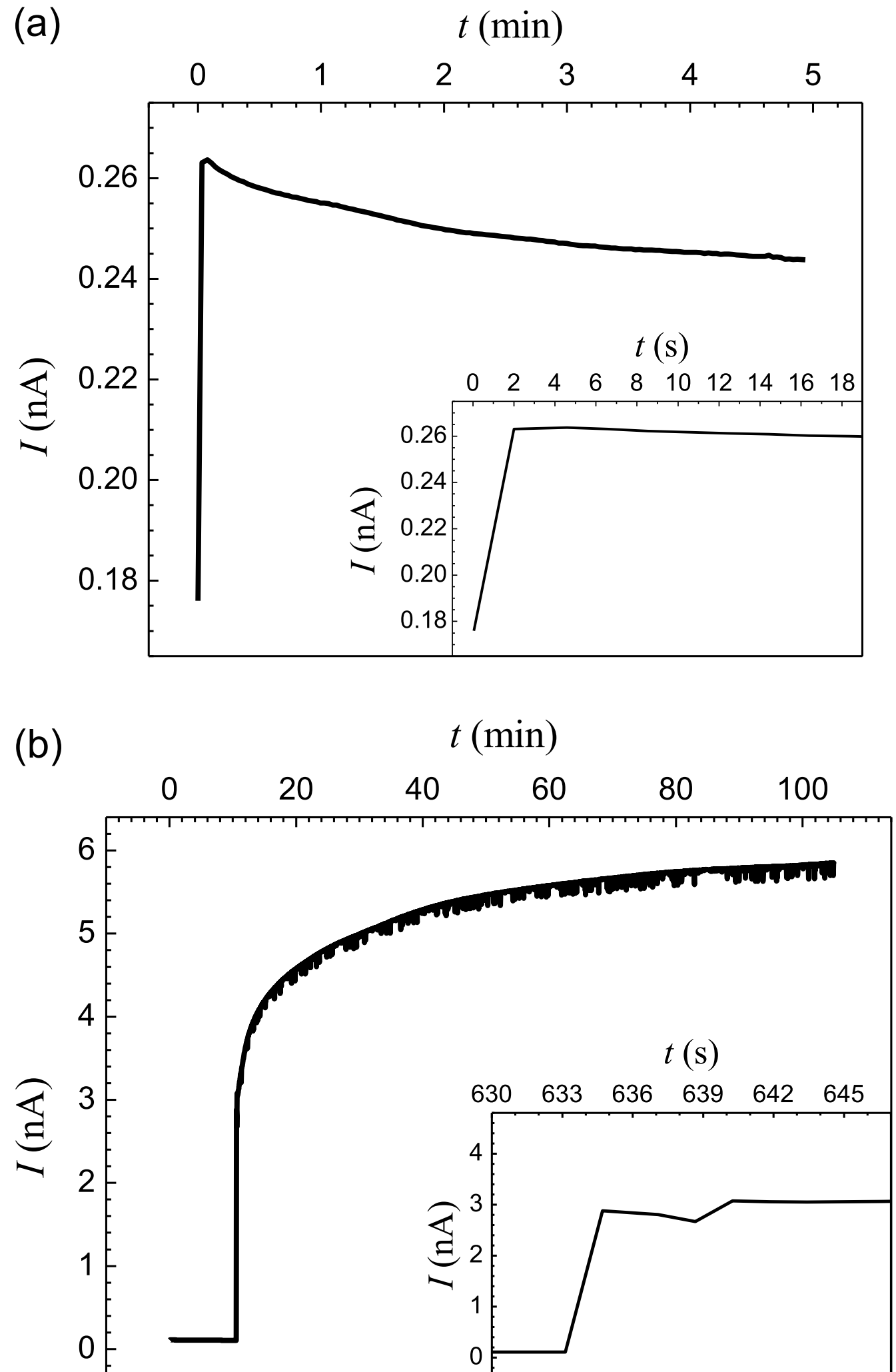

**Fig. 12.** Current transients of H1 device (a) under dark conditions upon applying 10 V at t = 0 min; (b) upon suddenly turning on the 740 nm LED light (with a photon flux of $10^{16}$ $cm^{-2}s^{-1}$) on a 10 V polarized sample. The insets correspond to a zoom of the initial part of the transients. In both cases, we observe an initial fast current-increasing transient of the order of seconds, followed by a much slower transient.

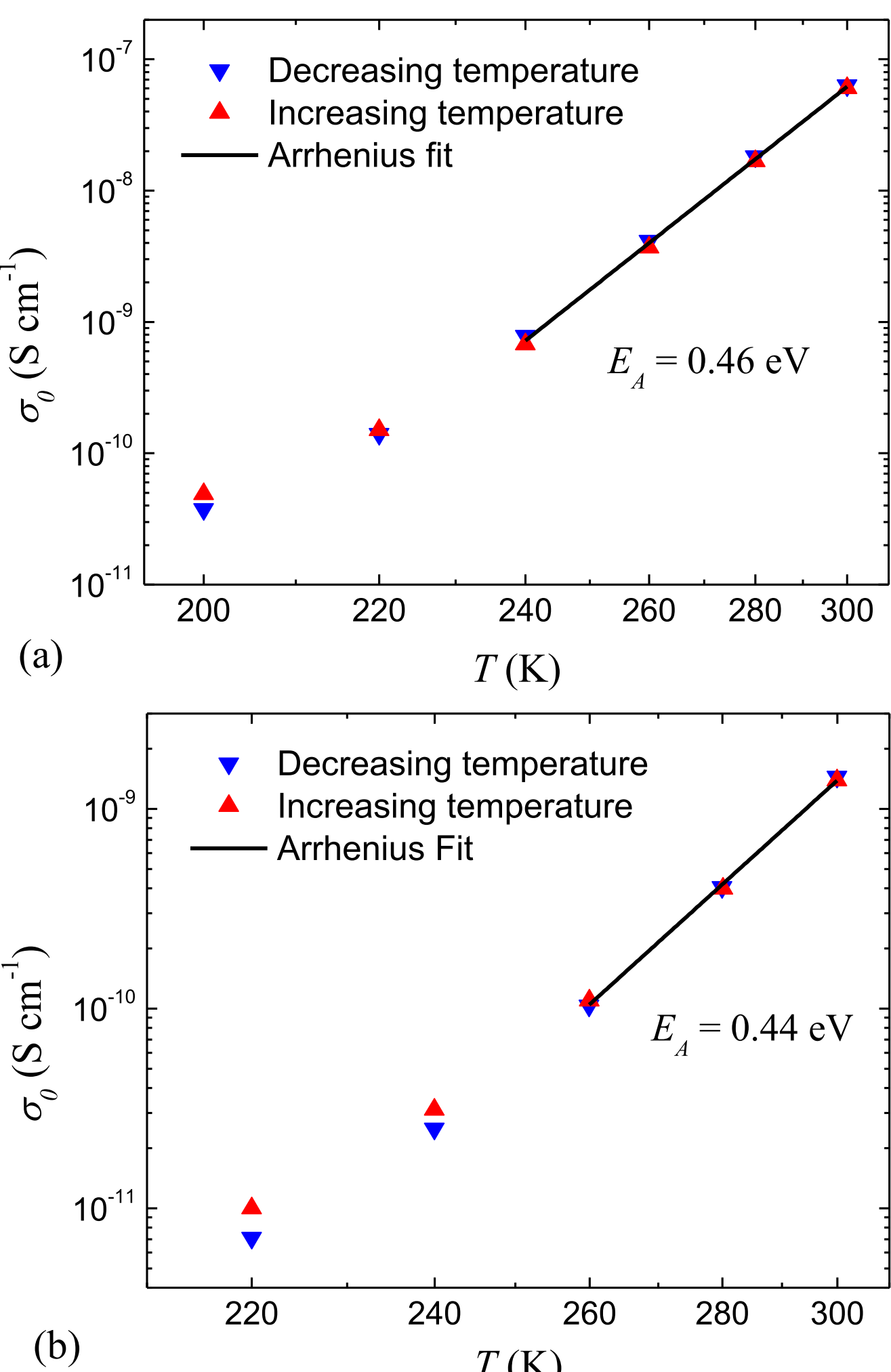

**Fig. 13.** Arrhenius plot of the quasi-steady-state dark conductivity for (a) H1 and (b) H2 devices. On the vertical and horizontal axes, we use logarithmic and reciprocal scales, respectively. Downward blue triangles and upward red triangles correspond to values extracted during decreasing and increasing temperature sweeps, respectively. At high temperatures, the conductivity shows an exponential dependence on temperature (appearing linear in this plot), indicative of free-carrier conduction. The activation energy corresponds to the position of the Fermi level relative to the nearest band.

b. Quasi-steady-state dark conductivity at different temperatures

The dark IV curves of the H1 device during temperature decrease and increase are shown in Figs. S5 and S6, respectively, while the corresponding curves for the H2 device are presented in Figs. S11 and S12. We observe a homogeneous linear response of the current with the voltage, indicating operation within the ohmic region. The conductivity is extracted from the slope of the linear fits.

Fig. 13 presents the extracted quasi-steady-state conductivity under dark, $\sigma_0$, for devices (a) H1 and (b) H2. As expected, we observe a monotonous increase in the dark conductivity with temperature. At high temperatures, we observe an exponential dependence, indicating extended-states conduction,

$$\sigma_0 = q\mu_M N_M \exp\left(\frac{-E_A}{k_b T}\right), \tag{7}$$

where $E_A$ is the Fermi level distance to the nearest band, $\mu_M$ and $N_M$ are the mobility and effective density of states for the majority carrier. The solid lines in Fig. 13 correspond to the fit of the highest temperature points with Eq. (7). Note that we use logarithmic and reciprocal scales in the vertical and horizontal axes, respectively. Assuming $N_M = 2 \times 10^{18}$ $cm^{-3}$ [32], we estimate from the fits that $\mu_M \sim 10$ $cm^2V^{-1}s^{-1}$ and $E_A \sim 0.46$ eV for the first device, and $\mu_M \sim 0.14$ $cm^2V^{-1}s^{-1}$ and $E_A \sim 0.44$ eV for the second one. Fitting the $\mu_M$ and $E_A$ values with the Meyer-Neldel formula [36], we obtain a Meyer-Neldel energy $E_{MN}$ of 5 meV. The Meyer-Neldel rule in the dark conductivity of thin films is typically related to the linear variation of the band gap and/or the Fermi level with temperature [50]. A large-scale study of the dark conductivity in halide perovskites is currently lacking in the literature; however, high-temperature activation energies of the dark conductivity, typically ranging from 0.4 to 0.5 eV, are characteristic of lead halide perovskites [25,51–53].

c. Quasi-steady-state photoconductivity (SSPC) at different temperatures

Figs. S7 and S8 show the IV curves of device H1 measured during temperature decrease under high and low light fluxes, respectively. Figs. S9 and S10 present the corresponding IV curves measured during temperature increase.

A linear region is observed for negative and low positive voltages. At high positive voltages, especially at elevated temperatures, a super-linear region appears, indicating asymmetry in the sample. Fig. S4 shows room temperature IV curves of the same device (H1) in both scan directions, demonstrating that the current response is independent of the scan direction. As usual, the conductivity under illumination (or photoconductivity) is estimated from the linear region.

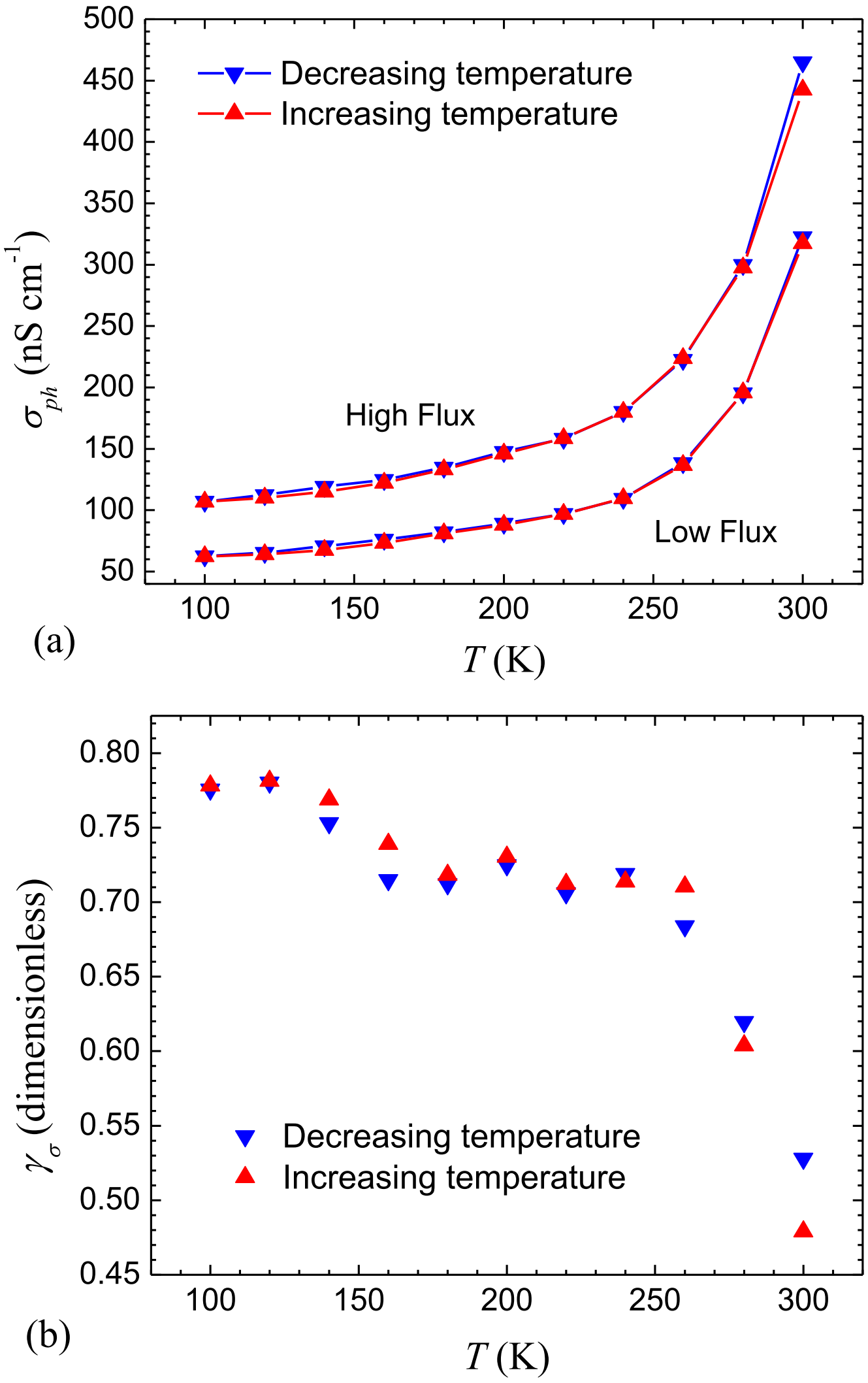

**Fig. 14.** (a) Conductivity of the H1 device under two generation rates ($10^{20}$ and $5 \times 10^{19}$ cm$^{-3}$s$^{-1}$) as a function of temperature. (b) Logarithmic derivative of the conductivity with respect to the generation rate for the H1 device versus temperature. Two successive temperature sweeps were performed, with data collected during the initial decreasing sweep (downward blue triangles) followed by the increasing sweep (upwards red triangles).

Fig. 14a shows the conductivity values of the H1 device at high and low generation rates ($10^{20}$ and $5 \times 10^{19}$ cm$^{-3}$s$^{-1}$, respectively), measured during both cooling and heating. No thermal hysteresis is observed, except at 300 K under high flux, indicating negligible material degradation during the measurements. Fig. 14b shows the logarithmic derivative $\gamma_\sigma = \Delta\left[\ln\left(\sigma_{ph}\right)\right]/\Delta[\ln(G_r)]$ extracted from the data in Fig. 14a.

Figs. S13 and S14 show the IV curves of device H2 measured during temperature decrease and increase, respectively. Panels (a) and (b) display the IV curves obtained under high and low light fluxes, respectively. The conductivity is extracted from a linear fit of the experimental data between 0 and 5 V, where the best linearity is observed.

Fig. S15a shows the conductivity values of the H2 device at high and low generation rates ($10^{20}$ and $5 \times 10^{19}$ cm$^{-3}$s$^{-1}$, respectively), measured during both cooling and heating. Fig. S15b shows the logarithmic derivative $\gamma_\sigma = \Delta\left[\ln\left(\sigma_{ph}\right)\right]/\Delta[\ln(G_r)]$ derived from the data in Fig. S15a. We observe thermal hysteresis at the three highest temperatures, indicating some degradation of the perovskite during the measurements [54]. In this case, the measurements were performed more than 10 times slower than in the H1 device. A higher degradation of the material is reasonable because the sample has been stressed for a significantly longer time.

The monotonic increase in photoconductivity (or photocurrent) with temperature, as shown in Fig. 14a, is a typical behavior observed in multi-cation lead halide perovskites [25,28,53]. The low-temperature photoconductivity quenching (decrease of the photoconductivity with temperature) that we observe in Fig. S15a has only been reported for a $MAPbI_3$ perovskite [55]. To our knowledge, this is the first report of this behavior for a mixed-cation mixed-halide perovskite. Using the zero-dimensional numerical simulation detailed in Section 5.1, we observe that the low-temperature photoconductivity quenching could be reproduced by adding an extra acceptor-like defect distribution very near to the conduction band edge or, as suggested in Ref. [55], by assuming mobilities that decrease with temperature.

$\gamma_\sigma$ has already been measured at similar generation rates for thin films of lead halide perovskites, primarily at room temperature [52,56] or within a much narrower temperature range [55]. The values reported in the literature range from 0.4 to 0.8, in good agreement with our measurements.

d. Quasi-steady-state photocarrier grating (SSPG) at room temperature

Fig. 15a presents the device H3 values of $\beta$ versus $k = 2\pi/\Lambda$ measured at different applied voltages with a photocarrier generation rate of $8 \times 10^{20}$ cm$^{-3}$s$^{-1}$. Practically, the same values of $\beta$ are obtained for the two lowest voltages; consequently, the RZW equation [Eq. (S10)] should be valid for electric fields equal to or lower than 60 Vcm$^{-1}$. We fit the experimental points obtained at 60 Vcm$^{-1}$ (solid line), as the signal to noise ratio is higher in this case, obtaining $L_D = 0.19 \pm 0.01$ μm and $\varphi_0 = 1.2$. Replacing the value of $\varphi_0$ in Eq. (S11) and assuming $\frac{\gamma_0^2}{1+G_2/G_1} = 0.85$ as usual [57], we obtain $\gamma_\sigma = 0.71$, which is in the range of expected values. The low-voltage SSPG technique has already been applied several times before to lead halide perovskite thin films, typically reporting $L_D$ values between 0.15 and 0.7 μm [24–26,28,51,52,56,58,59].

The quasi-steady-state photocurrent measured at different voltages while illuminating the sample with the main beam is presented in Fig. 15b. We waited a couple of seconds for stabilization at each voltage before measuring the current. Despite the super-linear dependence of the photocurrent with voltage ($J \propto V^{1.2}$), the data presented in Fig. 15a exhibit good linearity. Consistent with previous suggestions [60,61], the RZW equation remains valid in the non-linear or non-ohmic range of voltages.

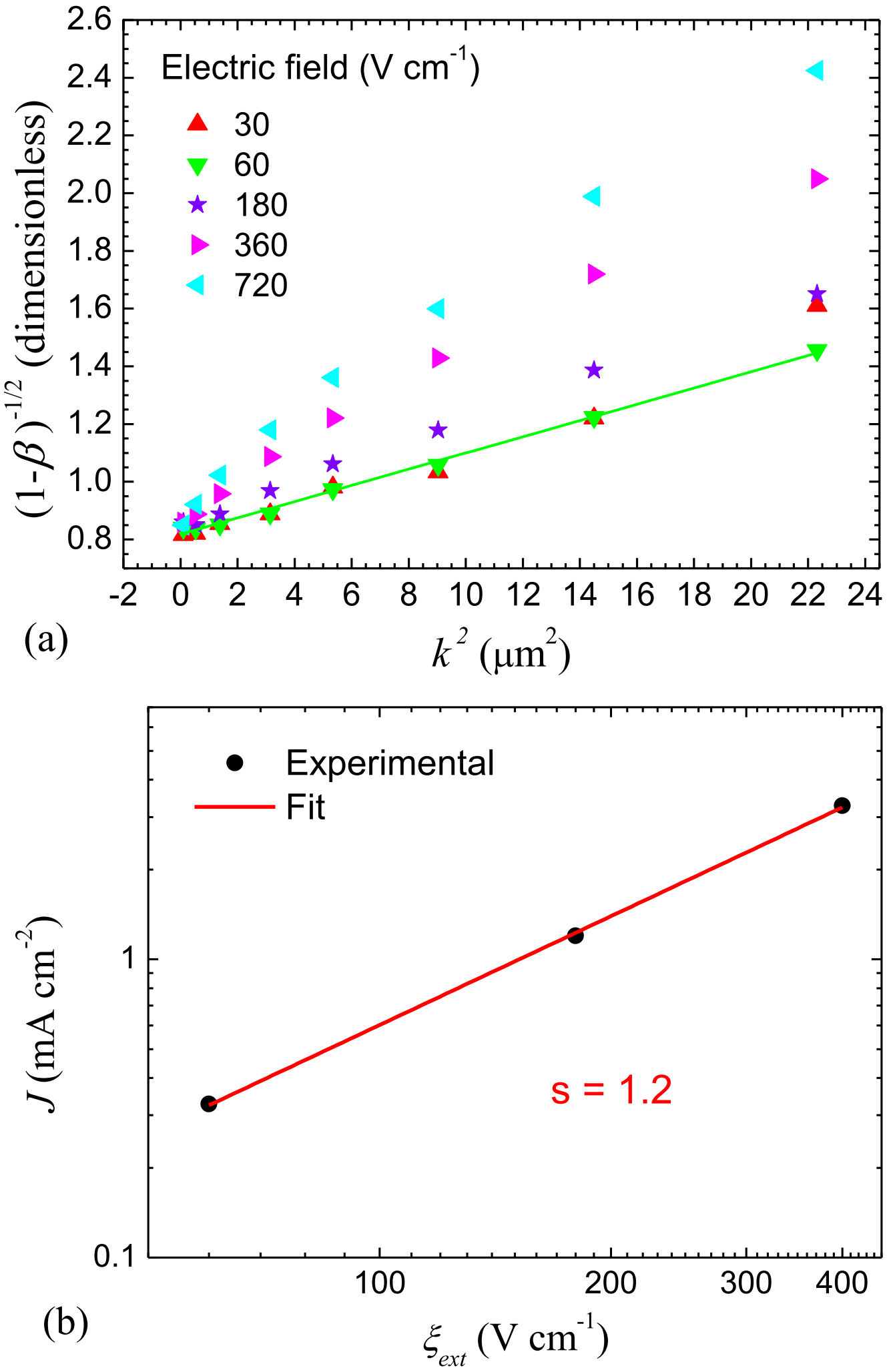

**Fig. 15.** (a) Steady-state photocarrier grating (SSPG) and (b) steady-state photocurrent (SSPC) measurements of the sample H3 at room temperature and under a generation rate of $8 \times 10^{20}$ cm$^{-3}$s$^{-1}$ using a 633 nm He-Ne laser. $\beta$ is the ratio of the AC signals generated when the beams have parallel and perpendicular polarizations. (a) $\beta$ versus $k = 2\pi/\Lambda$ measured at different applied voltages (symbols), where $\Lambda$ is the photocarrier grating period created by the interference of the beams. The solid line shows the linear fit of the data obtained with an applied electric field of 60 Vcm$^{-1}$. (b) Quasi-steady-state photocurrent measured at different voltages while illuminating the sample with the main beam.

## 5. Estimation of the density of states (DOS) and the material parameters

In this section, we propose the simplest perovskite density of defect states (DOS), along with the rest of the material parameters, which are consistent with the experimental results presented in the previous sections.

### 5.1. Lateral devices with coplanar contacts

For simulating the quasi-steady-state measurements performed in the lateral devices, we neglect the ionic migration, assume a uniform material and ideal ohmicity of the contacts. These assumptions and the uniform illumination, allow us to use a simpler zero-dimensional model, which has been extensively applied to explain the temperature dependence of SSPC and SSPG in thin film semiconductors (see Refs. [62,63], for instance). The equations of this numerical simulation are given in Section 5A of the SI.

Initially, we define a density of states within the bandgap (Eqs. S13-S15) and assign values to the material parameters within a reasonable range. At a fixed temperature, we extract the Fermi level by solving the dark neutrality equation (Eq. S(16)). At the average generation rate of SSPC measurements $G_r = 8 \times 10^{19}$ cm$^{-3}$s$^{-1}$, we solve simultaneously the continuity and charge neutrality equations (Eqs. S19 and S20), for extracting $\sigma_{ph}$ and $\gamma_\sigma$. We repeat this procedure for all the temperatures measured experimentally. At $T = 300$ K and the generation rate of SSPG experiment, $G_r = 8 \times 10^{20}$ cm$^{-3}$s$^{-1}$, we calculate $L_D$ by simultaneously solving the continuity and charge neutrality equations again. The material parameters are adjusted one by one until all the experimental results are reproduced.

The parameters that are kept fixed in this simulation are presented in Table 2. The room temperature effective density of states $N^0_{V,C}$ in lead halide perovskites is taken from Ref. [32]. The electrical bandgap of this material is estimated by fitting the solar cell IV measurements (see Section 5.2). We estimate the DOS at the band edges from $N(E_{V,C}) = (k_b 300)^{-1} N^0_{V,C}$, obtained by assuming the continuity of the DOS at the band edges with a negligible slope [50]. For the perovskite static permittivity, we use the same value that reproduces the experimental capacitance spectrum presented in Fig. 11a with $2 \times 10^{18}$ cm$^{-3}$ ions (this ionic concentration is estimated from the fit of the solar cell IV measurements presented in Section 5.2). Note that the ionic concentration and mobility are not relevant in quasi-steady-state measurements, since ideally no ionic migration occurs during these measurements.

The Urbach energy extracted from sub-bandgap optical absorption is equal to the highest band tail characteristic energy [50]. Above 100 K, we assume a homogeneous linear relationship between the characteristic energy of the band tails and the temperature, $E_{VBT,CBT} = \alpha_{V,C} k_b T$, because the Urbach energy exhibits a homogeneous linear relationship with temperature in this temperature range [19–23]. Assuming a maximum value of 28 meV for the Urbach energy at 300 K [19–23], we obtain $\alpha_{V,C} \leq 1.1$.

All estimates of room-temperature free-carrier mobilities in lead halide perovskite thin films fall within the range of 0.1-60 cm$^2$ V$^{-1}$ s$^{-1}$ [64,65]. For simplicity, we assume temperature-independent free carrier mobilities.

The reported capture coefficients of defect states at room temperature range from $10^{-13}$ to $10^{-6}$ cm$^3$ s$^{-1}$ [4,13,66,67]. For simplicity, we neglect their temperature dependence and assume equal electron and hole capture coefficients for each defect distribution [68].

The experimental values of $\sigma_{ph}$ and $\gamma_\sigma$ differ significantly between H1 and H2 samples (Fig. 14 and S15), despite having nearly the same high-temperature activation energy for the dark conductivity (Fig. 13). When attempting to find the parameters that simultaneously reproduce the experimental values of $\sigma_{ph}$ and $\gamma_\sigma$, we were unable to completely reproduce those of H2 device, particularly the sudden changes of $\gamma_\sigma$ with temperature (Fig. S15b). In this device, we performed the dark IV measurements separately from the IVs under illumination, performing each measurement more than ten times slower. It is possible that new defects were generated during these long dark IV measurements. In our zero-dimensional model, we assume that all defects are located in the bulk of the material; therefore, it is reasonable that we cannot reproduce the behavior produced by defects generated near the contacts. In this sample (H2), we waited 14 times more at each voltage before measuring the photocurrent. Therefore, the errors due to the slow transients (associated with defect migration) should be higher. Another possibility is that one of our previous assumptions, such as temperature-independent capture coefficients, is not valid in this sample [68]. From now on, we will only try to reproduce the $\sigma_{ph}$ and $\gamma_\sigma$ values of H1 device, which are simpler.

**Table 2**
Perovskite parameters that were kept fixed in the numerical simulation of the horizontal devices measurements.

| Parameter [unit] | Symbol | Value |
|---|---|---|
| Effective DOSs at 300 K [cm$^{-3}$] | $N^0_{V,C}$ | $2 \times 10^{18}$ |
| Electro bandgap energy [eV] | $E_g$ | 1.72 |
| DOS at the band edges [eV$^{-1}$ cm$^{-3}$] | $N(E_{V,C})$ | $7.7 \times 10^{19}$ |
| Relative permittivity [−] | $\varepsilon_r$ | 29 |

The extracted DOS, consistent with all previous experimental data, is sketched in Fig. 16. Donor-like distributions are green, and acceptor-like distributions are brown. The extracted parameters from the simulation of the lateral devices are listed in Table 3, along with the accepted range of values for each parameter. Note that $\alpha_{CBT}$ does not have any particular value assigned, which means that the results are independent of this parameter, as long as it remains in the accepted range of values. The acceptor and donor concentrations do not appear in Table 3 because $\sigma_0$, $E_F$, $\sigma_{ph}$, $\gamma_\sigma$ and $L_D$ only depend on the Effective Dopant Concentration (EDC) $N_{DON}$ - $N_{ACC}$, and not on their individual values.

The slopes of the band tails decrease linearly with temperature. Therefore, we present in Fig. 16 the band tails at the two extreme measured temperatures, 100 and 300 K. As usual, we assume donor-like (acceptor-like) character for the valence (conduction) band tail states [29,62,63]. The wider Acceptor-like Gaussian Distribution (AGD) centered at 1.51 eV completely dominates the recombination in the entire temperature range. The narrower Gaussians distributions located very near the band edges just correspond to completely ionize donors and acceptors. Their concentration is estimated from the simulation of the solar cell experiments presented in Section 5.2. They are practically equal in size because $N_{DON}$ - $N_{ACC}$ is several orders of magnitude lower than $N_{DON}$ and $N_{ACC}$. The blue and red vertical dashed lines correspond to the Fermi level at 100 and 300 K, respectively.

Let us compare our experimental results with those obtained from Fig. 16 DOS and the parameters of Tables 2 and 3. The activation energy of the dark conductivity ($E_C - E_F$), as extracted from the numerical simulation, is presented in Fig. S16. We observe an increase in the activation energy with temperature, indicating that the semiconductor becomes more intrinsic as the temperature rises. According to Fig. S16, the activation energy lies in the range of 0.44 to 0.48 eV between 230 and 300 K, which is very similar to the values extracted experimentally (0.44 and 0.46 eV) in the same temperature range (see Fig. 13).

**Table 3**
Parameters changed in the numerical simulation in order to fit the experimental data of the lateral devices. The third column corresponds to the allowed ranges of variation. The fourth column corresponds to the values that produce the best fit of the data.

| Parameter [unit] | Symbol | Range | Value |
|---|---|---|---|
| Electron mobility [cm$^2$ V$^{-1}$ s$^{-1}$] | $\mu_n$ | 0.1 - 60 | 55 |
| Hole mobility [cm$^2$ V$^{-1}$ s$^{-1}$] | $\mu_p$ | 0.1 - 60 | 0.1 |
| VBT constant [−] | $\alpha_V$ | 0.3 - 1.1 | 1.1 |
| VBT capture coefficients [cm$^3$ s$^{-1}$] | $c_{n,p}^{VBT}$ | $10^{-11}$ - $10^{-6}$ | $2 \times 10^{-7}$ |
| CBT constant [−] | $\alpha_C$ | 0.3 - 1.1 | - |
| CBT capture coefficients [cm$^3$ s$^{-1}$] | $c_{n,p}^{CBT}$ | $10^{-11}$ - $10^{-6}$ | $2 \times 10^{-7}$ |
| AGD position [eV] | $E_A$ | 0 - 1.72 | 1.51 |
| AGD stand. deviation [meV] | $\omega_A$ | 2 - 60 | 50 |
| AGD concentration [cm$^{-3}$] | $N_A^0$ | $10^{13}$ - $10^{19}$ | $1.2 \times 10^{18}$ |
| AGD capture coefficients [cm$^3$ s$^{-1}$] | $c_{n,p}^A$ | $10^{-13}$ - $10^{-6}$ | $5 \times 10^{-8}$ |
| EDC concentration [cm$^{-3}$] | $N_{DON}$ - $N_{ACC}$ | $10^{10}$ - $10^{16}$ | $2 \times 10^{14}$ |

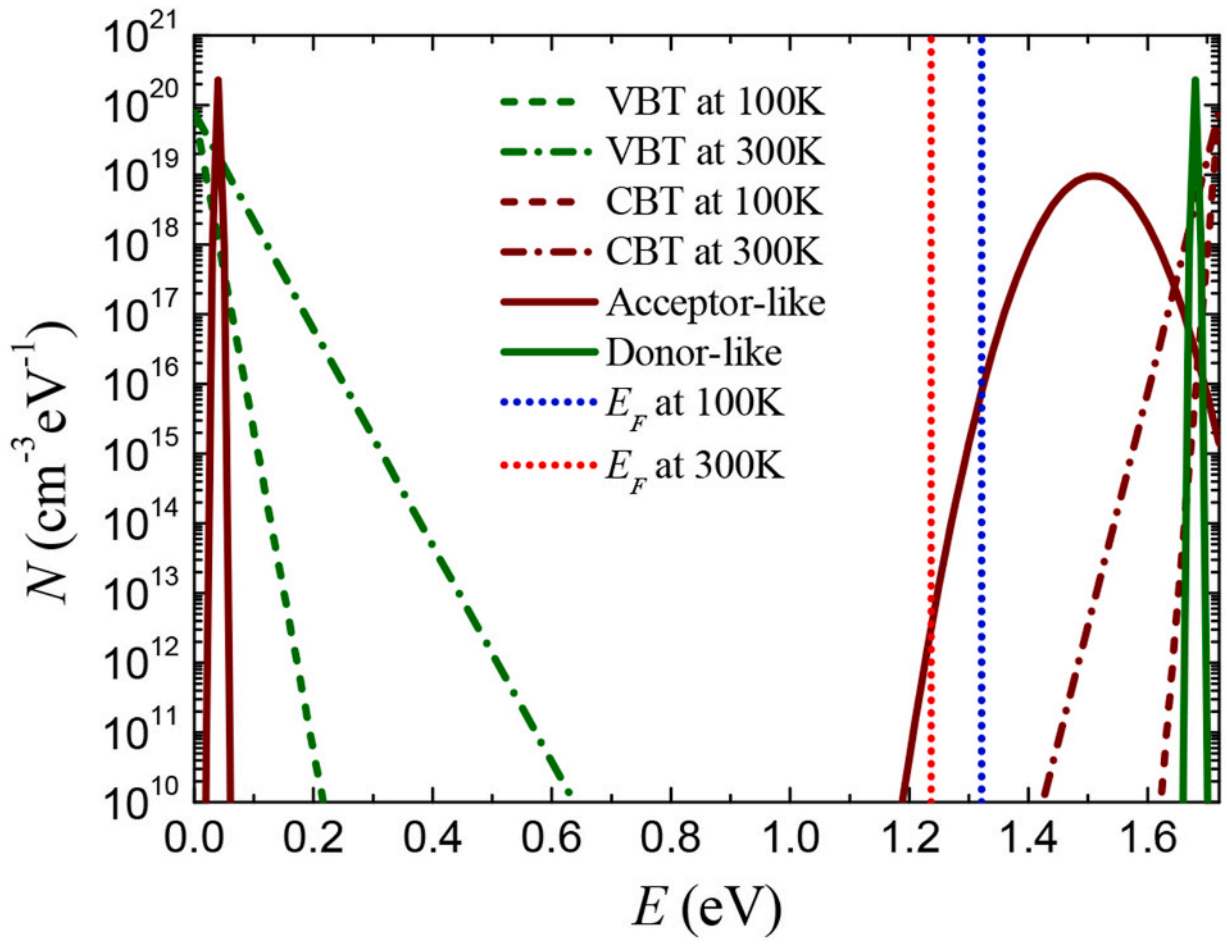

**Fig. 16.** Density of states (DOS) inside the bandgap that is consistent with all experimental data. The slope of each band tail increases linearly with temperature (with a proportionality constant $\alpha$). Valence and Conduction Band Tails (VBT and CBT) at the two extreme measured temperatures are shown. The plotted CBT corresponds to $\alpha_{CBT} = 0.5$, although the results are independent of this parameter as long as it is lower than $\alpha_{VBT}$. The narrower Gaussian distributions near the valence and conduction band edges represent the acceptor and donor concentrations, respectively. The position of the Fermi level ($E_F$) at the two extreme temperatures is also indicated. Parameter values for this DOS are listed in Tables 2 and 3.

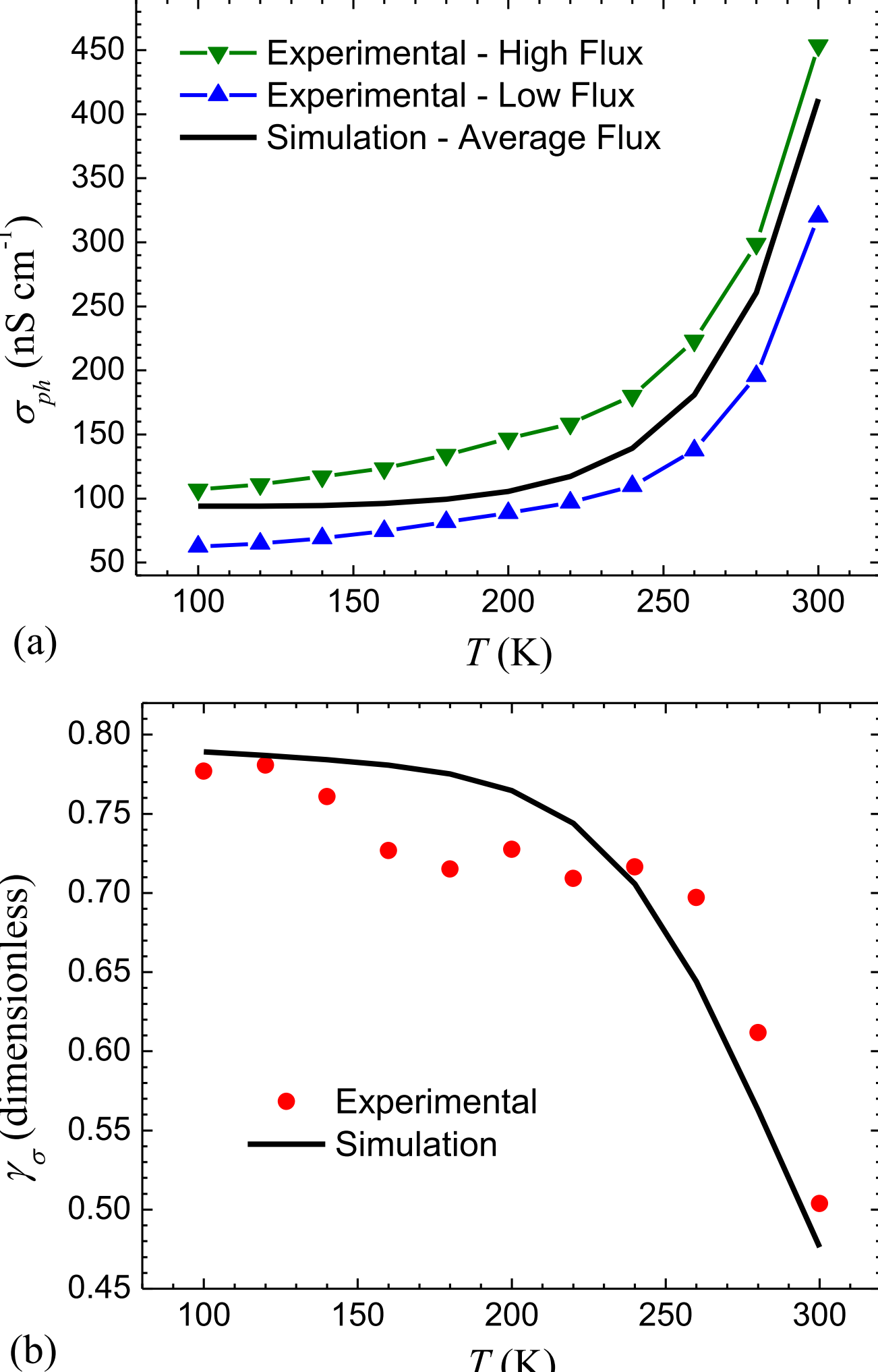

**Fig. 17.** (a) Conductivity versus temperature for the H1 device under two generation rates: $10^{20}$ cm$^{-3}$s$^{-1}$ (downward green triangles) and $5 \times 10^{19}$ cm$^{-3}$s$^{-1}$ (upward blue triangles). (b) Extracted logarithmic derivative of conductivity with respect to generation rate versus temperature for the H1 device (red circles). Solid lines correspond to the numerical simulation results at an intermediate generation rate of $8 \times 10^{19}$ cm$^{-3}$s$^{-1}$ using Fig. 16 DOS with the parameter values given in Tables 2 and 3.

The symbols in Fig. 17a correspond to the conductivity versus temperature of H1 device under the two generation rates ($10^{20}$ and $5 \times 10^{19}\,\mathrm{cm^{-3}s^{-1}}$), and in Fig. 17b to the logarithmic derivative of the conductivity with respect to the generation rate calculated from the previous values. These symbols correspond to the average of the values obtained during the decrease and increase of temperature presented in Fig. 14. The solid lines correspond to the values obtained from the numerical simulation with Tables 2 and 3 parameters and Fig. 16 DOS, at an intermediate generation rate of $8 \times 10^{19}\,\mathrm{cm^{-3}s^{-1}}$. We observe a very good agreement between the experimental and the simulated values.

Fig. 18a presents again the $\sigma_{ph}$ extracted from the numerical simulation, but detailing the contribution of electrons, $\sigma_n = q\mu_n n$, and holes, $\sigma_p = q\mu_p p$, to the total value. Similarly, Fig. 18b presents the $\gamma_\sigma$ extracted from the numerical simulation with the logarithmic derivatives of $n$ and $p$ with respect to $G_r$ ($\gamma_n$ and $\gamma_p$, respectively). Note that $n$ increases and $\gamma_n$ decreases with temperature because the recombination center is near the conduction band, whereas $p$ and $\gamma_p$ remain constant. As the temperature increases, more trapped electrons in the AGD are emitted to the conduction band, increasing the free-electron concentration $n$. $\gamma_n$ decreases with temperature because at higher temperatures, there are fewer trapped electrons in the AGD to be optically excited.

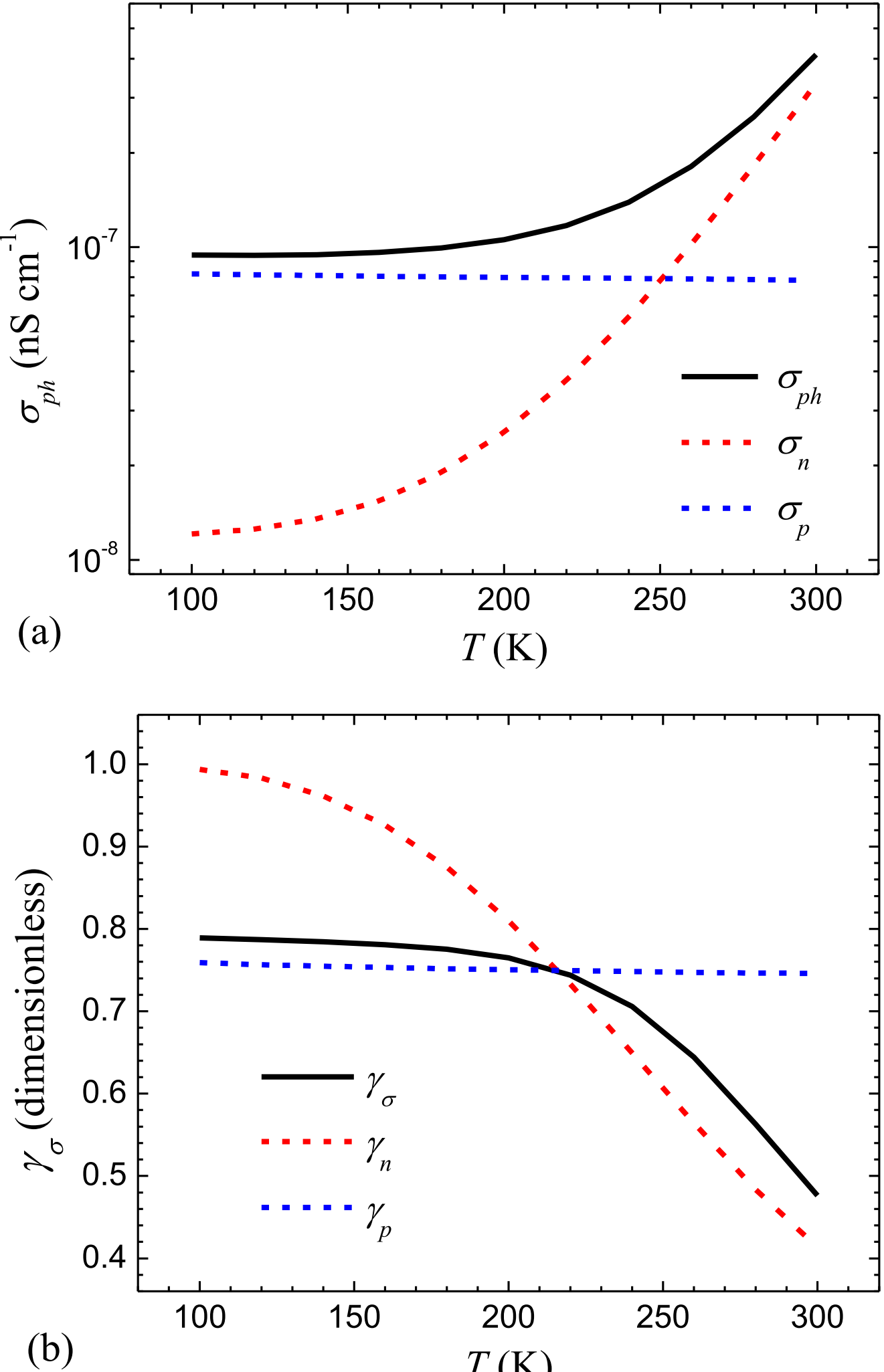

**Fig. 18.** Numerical simulation results under a generation rate of $8 \times 10^{19}\,\mathrm{cm^{-3}s^{-1}}$ using Fig. 16 DOS with the parameter values of Tables 2 and 3. Panel (a) shows the photoconductivity versus temperature (solid line) and its separation into electron ($\sigma_n = q\mu_n n$) and hole components ($\sigma_p = q\mu_p p$), represented by dashed lines. Panel (b) shows the logarithmic derivative of conductivity with respect to generation rate versus temperature (solid line), with the logarithmic derivatives of the electron ($\gamma_n$) and hole ($\gamma_p$) concentrations with respect to generation rate (dashed lines).

At low temperatures, the photoconductivity is practically dominated by holes, i.e., $\sigma_{ph} \sim q\mu_p p$ and $\gamma_\sigma \sim \gamma_p$. As the temperature increases, the electron contribution increases and at about 250 K, equates the hole contribution. Above 180 K, the electron contribution is not negligible anymore, causing the increase in $\sigma_{ph}$ and the decrease of $\gamma_\sigma$ with temperature.

From the numerical simulation with Fig. 16 DOS and the parameters in Tables 2 and 3, we extract $L_D = 0.18\,\mu\mathrm{m}$ at $G_r = 8 \times 10^{20}\,\mathrm{cm^{-3}s^{-1}}$ and $T = 300\,\mathrm{K}$, which (within the experimental error) matches the value obtained from SSPG under the same conditions. Note that the calculated ambipolar diffusion length is lower, $L_a = 0.10\,\mu\mathrm{m}$, indicating that the second and third terms of Eq. (S12) are not negligible in this case.

The presence of just a dominant recombination center at about 0.21 eV from the conduction band is in agreement with the results reported by Azulay et al. [28] for a similar perovskite [$FA_{0.79}MA_{0.16}Cs_{0.05}Pb(I_{0.83}Br_{0.17})_3$]. Using scanning tunneling spectroscopy, they only observed a defect state at about 0.25 eV below the conduction band. They also perform SSPC and SSPG measurements at different temperatures and observe that $q\mu_n n_0$ increases with temperature, while $q\mu_p p$ remains practically constant. Indicating that the recombination centers should be at about 0.2-0.3 eV from the conduction band.

All our experimental data is also consistent with the symmetrical DOS and the parameters obtained by changing the electron by the hole parameters, and vice versa. These new parameters and DOS are provided in Table S1 and Fig. S17, respectively. In this case, the semiconductor is p-type at room temperature, and the recombination center is a Donor-like Gaussian Distribution (DGD) located 0.21 eV above the valence band edge, which is not consistent with Ref. [28] results.

### 5.2. Solar cells

Let us compare the results obtained from thermal admittance spectroscopy with Fig. 16 DOS and Tables 2 and 3 parameters. According to the simulations presented in Fig. 11a, we need at least $\sim 10^{17}$ mobile charged species to reproduce the experimental capacitance step. The recombination centers consist of acceptor-like states predominantly located above the Fermi level. Consequently, of the total defect density ($1.2 \times 10^{18}\,\mathrm{cm^{-3}}$), approximately $2 \times 10^{14}\,\mathrm{cm^{-3}}$ are charged at thermal equilibrium. Under the same conditions, charged defects in the band tails are even less abundant and remain below $2 \times 10^{10}\,\mathrm{cm^{-3}}$ over the measured temperature range. Therefore, these charged species are not enough to produce the low-frequency capacitance steps of Fig. 8a and b. The only viable candidates of the extracted DOS are the donors and/or acceptors. Donors likely correspond to halide vacancies $V_X^+$, especially iodine vacancies, which are typically identified as the most mobile species [27,33,69,70]. Theoretical studies predict activation energies between 0.08 and 0.59 eV for this species [69], which is consistent with our TAS measurements. Finally, note that in the extracted density of states, there are no defects near mid-gap that could originate extra capacitance steps.

Let us evaluate if the defect concentration of Fig. 16 with Tables 2 and 3 parameters and the high concentration of mobile dopants ($>10^{17}\,\mathrm{cm^{-3}}$) extracted from the capacitance spectra are consistent with the steady-state IV measurements presented in Fig. 6. For simulating the IVs, we simplify the DOS of Fig. 16 by neglecting the band tails and concentrating the recombination centers in the center of the Gaussian (0.21 eV below the conduction band edge). This approximation is justified because the recombination is completely dominated by this Gaussian distribution. The other parameters of the device that are not included in Tables 2 and 3 correspond to those used in the capacitance simulation presented in Fig. 11, with the exception of the electron and hole mobilities of the transport layers that are higher in this simulation (0.1 $cm^2 V^{-1} s^{-1}$). For simplicity, we assume that the generation rate is uniformly localized in the perovskite layer.

Fig. 19 shows the best fits obtained from the numerical simulation (symbols) to the experimental IVs of sample S5 (lines) using the same ionic concentrations used for the simulation of the capacitance spectrum. They are obtained with a uniform generation rate of $2.5 \times 10^{21}$ $cm^{-3}s^{-1}$ in the perovskite layer. We clearly observe that the sudden increase of the current with the voltage is better reproduced with the higher ionic concentration. By fitting the IVs curves, we obtain a better estimation of the mobile ionic parameters. Some of the device parameters used in the simulation of capacitance spectrum (see Fig. 11) were modified in order to fit the IV curves: the perovskite bandgap, recombination parameters, and transport layers mobilities. However, these parameters do not have any influence on the capacitance spectra. Therefore, we obtain the same capacitance curves of Fig. 11 with these new values of the parameters. Table 4 summarizes the material parameters estimated from the fitting of the solar cell experiments that are not included in Tables 2 and 3.

In the horizontal devices, the perovskite is surrounded by regular Corning glass and $Al_2O_3$, whereas in the solar cells, it is surrounded by TaTm and C60. Therefore, some difference in the defect landscape could be expected [25,41]. Despite this warning, we have successfully reproduced the complete set of measurements using the unique set of parameters, listed in Tables 2–4. For practicality, this complete list of parameters is given in Table S2.

## 6. Conclusion

This study presents a comprehensive optoelectronic characterization of vacuum-deposited $FA_{0.7}Cs_{0.3}Pb(I_{0.9}Br_{0.1})_3$ perovskite thin films in both vertical and lateral device configurations. Standard current-voltage (IV) and thermal admittance spectroscopy (TAS) measurements in solar cells are combined with lateral current measurements in perovskite films with top coplanar electrodes. From the lateral devices, we extract material properties at different temperatures, including the dark conductivity, photoconductivity and its generation-rate derivative. Using the SSPG technique, we also estimate the room temperature ambipolar diffusion length of photocarriers.

We start by reviewing TAS in perovskite solar cells. A one dimensional (1D) numerical simulation reveals that the experimentally observed low-frequency capacitance step in short-circuited devices can arise from either defects with high activation energies (>0.6 eV) or ionic species with low mobility ($<10^{-6}$ $cm^2 V^{-1} s^{-1}$) in the perovskite layer. We identify inconsistencies between typical analytical formulas and standard drift-diffusion models, leading to the development of new formulas. From the high-frequency capacitance plateau, we estimate a lower bound for the perovskite relative permittivity, $\varepsilon_r > 22$. The low-frequency capacitance steps shift exponentially to higher frequencies with increasing temperature; their low activation energies (~ 0.34 eV) exclude free carrier trapping and emission by defects as their cause, pointing instead to ionic migration.

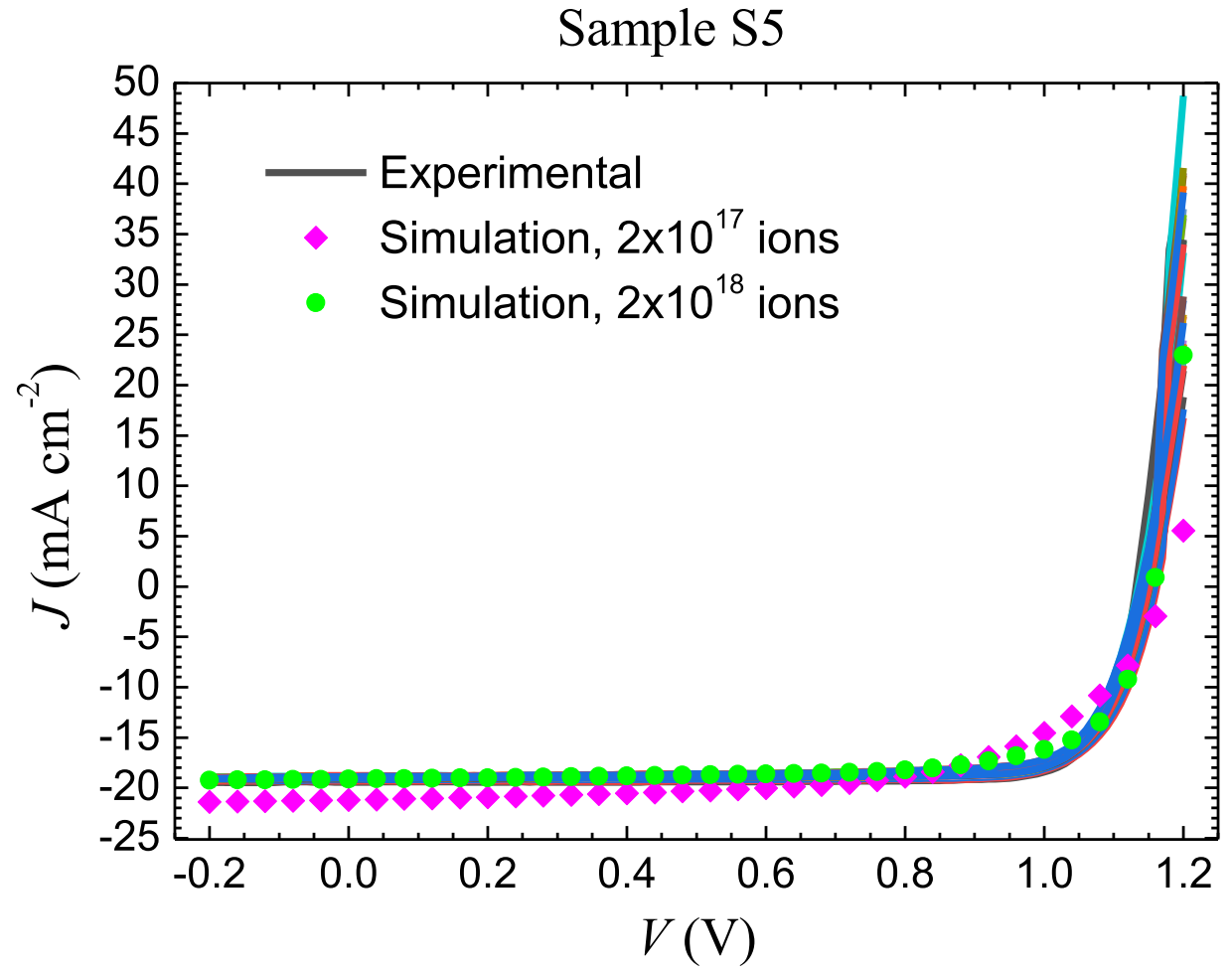

**Fig. 19.** JV curves of the 16 solar cells of samples S5 (lines) and the fits obtained with the numerical simulation (symbols).

**Table 4**
Parameter values of the perovskite (PVK) and transport layers (TLs) that reproduce the dark capacitance vs. frequency of the short-circuited solar cells, and their IV curves under standard conditions when assuming a uniform generation rate of $2.5 \times 10^{21}$ $cm^{-3}s^{-1}$ in the perovskite layer.

| Parameter [unit] | Symbol | Value |
|---|---|---|
| PVK donor concentration [$cm^{-3}$] | $N_{DON}$ | $2 \times 10^{18}$ |
| PVK donor mobility at 300K [$cm^2 V^{-1} s^{-1}$] | $\mu_{DON}^{300K}$ | $1 \times 10^{-8}$ |
| Average activation energy of the PVK donor mobility [eV] | $E_{\mu_{DON}}$ | 0.34 |
| HTL relative permittivity [–] | $\varepsilon_r^{HTL}$ | 6 |
| ETL relative permittivity [–] | $\varepsilon_r^{ETL}$ | 5 |
| Free-carrier mobilities of the TLs [$cm^2 V^{-1} s^{-1}$] | $\mu_{n,p}^{HTL,ETL}$ | 0.1 |

In the lateral (or horizontal) devices, we observe an exponential increase in the lateral dark conductivity with temperature above 240 K, indicating free carrier conduction in extended states. From the Arrhenius plot, we extract activation energies of 0.44-0.46 eV, corresponding to the Fermi level position relative to the nearest band edge. At a generation rate of $\sim 8 \times 10^{19}$ $cm^{-3}s^{-1}$ and increasing temperature, we observe that the lateral photoconductivity increases, while its generation-rate logarithmic derivative decreases. At a generation rate of $\sim 8 \times 10^{20}$ $cm^{-3}s^{-1}$ and room temperature, we estimate a SSPG diffusion length of $0.19 \pm 0.01$ μm.

The experimental results in the horizontal devices are reproduced using a zero-dimensional numerical simulation, which allow us to estimate the PVK free-carrier mobilities and the defect distribution inside the bandgap, including their capture coefficients. The defect distribution consists of: (1) exponential band tails whose slopes decrease linearly with temperature; (2) a Gaussian distribution (with $1.2 \times 10^{18}$ $cm^{-3}$ defects) centered 0.21 eV from the band edge, which completely dominates the recombination; and (3) an effective dopant concentration $|N_{DON} - N_{ACC}|$ of $2 \times 10^{14}$ $cm^{-3}$ that compensates the charge of the recombination centers.

The only charged species of the extracted DOS sufficient to produce the low-frequency capacitance steps are the dopants (acceptors and/or donors), indicating that the mobile ionic species correspond to dopants rather than recombination centers. The remaining material parameters are estimated by fitting the TAS and standard IV measurements of the solar cells with the 1D simulation. We estimate the perovskite electrical bandgap (1.72 eV), its mobile dopant concentration ($2 \times 10^{18}$ $cm^{-3}$) and the free carrier mobilities of the transport layers from the IV measurements. The average activation energy of the dopant mobility (0.34 eV), its 300 K average value ($10^{-8}$ $cm^2 V^{-1} s^{-1}$), and the relative permittivities of the PVK (29), HTL (6), and ETL (5) are estimated from TAS measurements.

### Declaration of generative AI and AI-assisted technologies in the manuscript preparation process

During the preparation of this work the authors used AI-assisted technologies in order to discuss the scientific literature. The authors take full responsibility for the content of the published article.

### CRediT authorship contribution statement

**L. Kopprio:** Conceptualization, Data curation, Formal analysis, Investigation, Methodology, Software, Validation, Visualization, Writing – original draft. **J. Caram:** Investigation, Writing – review &

editing. **S. Le Gall:** Conceptualization, Investigation, Project administration, Supervision, Validation, Writing – review & editing. **F. Ventosinos:** Conceptualization, Investigation, Writing – review & editing. **L. Gil-Escrig:** Resources, Writing – review & editing. **H.J. Bolink:** Resources, Writing – review & editing. **J. Alvarez:** Funding acquisition, Resources, Writing – review & editing. **C. Longeaud:** Investigation, Writing – review & editing. **J.-P. Kleider:** Conceptualization, Formal analysis, Methodology, Supervision, Writing – review & editing. **J. Schmidt:** Funding acquisition, Project administration, Supervision, Writing – review & editing.

## Declaration of competing interest

The authors declare that they have no known competing financial interests or personal relationships that could have appeared to influence the work reported in this paper.

## Acknowledgments

The authors thank Mateo Tentor (IFIS-Litoral), James Connolly (GeePs), Olivier Hubert (GeePs), Arthur Julien (IPVF), and especially Juan Pablo Medina (TOTAL), Pilar Lopez-Varo (IPVF) and Jonathan Parion (IMEC) for the fruitful discussions and constructive remarks. The authors acknowledge financial support from ECOS-SUD/MINCYT project (A22E02) and TRIUMPH Horizon Europe project (101075725).

## Appendix A. Supplementary data

Supplementary data to this article can be found online at https://doi.org/10.1016/j.solmat.2026.114557.

## Data availability

Data will be made available on request.